\documentclass[journal]{vgtc}                     %

\title{Comparative Evaluation of Animated Scatter Plot Transitions}

\author{%
  \authororcid{Nils Rodrigues}{0000-0002-1485-8249},
  \authororcid{Frederik L. Dennig}{0000-0003-1116-8450},
  Vincent Brandt,
  \authororcid{Daniel A. Keim}{0000-0001-7966-9740}, and
  \authororcid{Daniel Weiskopf}{0000-0003-1174-1026}
}

\authorfooter{
  \item
  	Nils Rodrigues and Daniel Weiskopf are with Visualization Research Center (VISUS), University of Stuttgart.
  	E-mail: {firstname}.{lastname}@visus.uni-stuttgart.de.
    
  \item
  	Frederik Dennig and Daniel A. Keim are with University of Konstanz.
  	E-mail: {firstname}.{lastname}@uni-konstanz.de.

  \item Vincent Brandt is with University of Stuttgart.
  	E-mail: st161848@stud.uni-stuttgart.de.
}

\abstract{%
  Scatter plots are popular for displaying 2D data, but in practice, many data sets have more than two dimensions.
For the analysis of such multivariate data, it is often necessary to switch between scatter plots of different dimension pairs, e.g., in a scatter plot matrix (SPLOM).
Alternative approaches include a \enquote{grand tour} for an overview of the entire data set or creating artificial axes from dimensionality reduction (DR).
A cross-cutting concern in all techniques is the ability of viewers to find correspondence between data points in different views.
Previous work proposed animations to preserve the mental map between view changes and to trace points as well as clusters between scatter plots of the same underlying data set.
In this paper, we evaluate a variety of spline- and rotation-based view transitions in a crowdsourced user study focusing on ecological validity.
Using the study results, we assess each animation's suitability for tracing points and clusters across view changes.
We evaluate whether the order of horizontal and vertical rotation is relevant for task accuracy.
The results show that rotations with an orthographic camera or staged expansion of a depth axis significantly outperform all other animation techniques for the traceability of individual points.
Further, we provide a ranking of the animated transition techniques for traceability of individual points.
However, we could not find any significant differences for the traceability of clusters.
Furthermore, we identified differences by animation direction that could guide further studies to determine potential confounds for these differences.
We publish the study data for reuse and provide the animation framework as a D3.js plug-in.
}


\teaser{
  \centering
  \includegraphics[width=0.8\linewidth]{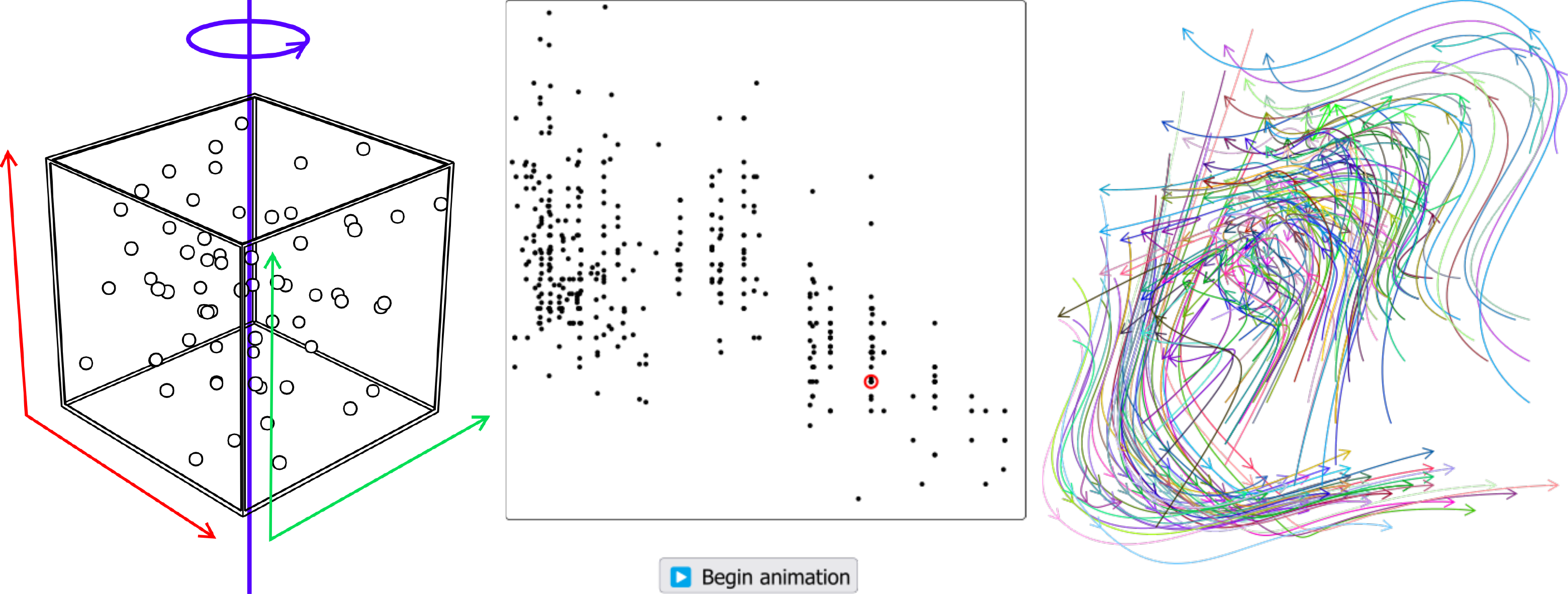}
  \caption{%
    We implemented six animation techniques to transition between scatter plots of different dimensions from the same underlying data set.
    The movement of data points during the animations is either based on rotation of 3D cubes (left) or splines (right).
    To evaluate the animated transitions, we conducted a crowdsourced user study where participants had to trace individual dots (center) or clusters.
  }
  \label{fig:teaser}
}

\graphicspath{{figures/}{./}} %

\usepackage{tabu}                      %
\usepackage{booktabs}                  %
\usepackage{lipsum}                    %
\usepackage{mwe}                       %

\usepackage{mathptmx}                  %

\PassOptionsToPackage{table}{xcolor}

\AtEndPreamble{%

\hypersetup{
    hidelinks
}

\captionsetup[sub]{font={scriptsize,sf}}

\usepackage[nameinlink]{cleveref}
\crefname{figure}{Fig.}{Figs.}
\Crefname{figure}{Figure}{Figures}
\crefname{section}{Sec.}{Secs.}
\Crefname{section}{Section}{Sections}
\crefname{table}{Table}{Tables}
\Crefname{table}{Table}{Tables}

\usepackage{amssymb}
\DeclareRobustCommand{\Chi}{\raisebox{0.3ex}{$\chi$}}

\usepackage{csquotes}

\usepackage{xifthen}

\usepackage{mfirstuc}

\usepackage[upgrade]{acro}
\acsetup{
    make-links=false,
    list/display=all,
    pages/display=none
}

\DeclareAcronym{SP}{
  short=SP,
  long=scatter plot
}
\DeclareAcronym{SPLOM}{
  short=SPLOM,
  long=scatter plot matrix,
  long-plural-form=scatter plot matrices
}
\DeclareAcronym{DR}{
  short=DR,
  long=dimensionality reduction
}

\def\TablerefTabPoints{Tables 1 and 2}
\def\TablerefTabPointsAlpha{Table 1}

\def\TablerefTabClusters{Table 3}
\def\TablerefTabDirection{Table 4}
\def\TablerefTabSubjectivePointsEase{Table 5}
\def\TablerefTabSubjectivePointsFrequent{Table 6}
\def\TablerefTabSubjectivePointsFast{Table 7}

\newcommand{\inlineIcon}[2][]{\raisebox{0pt}{{%
  \ifthenelse{\isempty{#1}}{%
    \def\iconheight{0.7em}%
  }{%
    \def\iconheight{#1}%
  }%
  \includegraphics[height=\iconheight,width=1.5em,keepaspectratio]{icons/#2.pdf}}%
}}

\newcommand{\straightIcon}{\inlineIcon{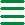}}
\newcommand{\bundledIcon}{\inlineIcon{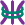}}
\newcommand{\timeoffsetIcon}{\inlineIcon{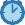}}
\newcommand{\stagedIcon}{\inlineIcon{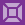}}
\newcommand{\perspectiveIcon}{\inlineIcon{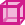}}
\newcommand{\orthographicIcon}{\inlineIcon{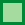}}

\newcommand{\straightAbbrev}[0]{STR}
\newcommand{\bundledAbbrev}[0]{BUN}
\newcommand{\timeoffsetAbbrev}[0]{TIM}
\newcommand{\stagedAbbrev}[0]{STA}
\newcommand{\perspectiveAbbrev}[0]{PER}
\newcommand{\orthographicAbbrev}[0]{ORT}
\newcommand{\straight}[0]{\straightIcon~\straightAbbrev}
\newcommand{\bundled}[0]{\bundledIcon~\bundledAbbrev}
\newcommand{\timeoffset}[0]{\timeoffsetIcon~\timeoffsetAbbrev}
\newcommand{\staged}[0]{\stagedIcon~\stagedAbbrev}
\newcommand{\perspective}[0]{\perspectiveIcon~\perspectiveAbbrev}
\newcommand{\orthographic}[0]{\orthographicIcon~\orthographicAbbrev}
\newcommand{\hypPoint}[0]{\textbf{HP}}
\newcommand{\hypCluster}[0]{\textbf{HC}}
\newcommand{\hypDirection}[0]{\textbf{HD}}
\newcommand{\hypRating}[0]{\textbf{HR}}
\newcommand{\hf}[0]{\textbf{hf}}
\newcommand{\vf}[0]{\textbf{vf}}
\newcommand{\bo}[0]{\textbf{bo}}
\newcommand{\rank}[1]{\textbf{#1.}}

\DeclareRobustCommand{\timeScale}{%
    \tikz{%
        \node[rectangle, left color = blue, right color = red]{\;\;\;\;\;\;\;};%
    }%
}

\newcommand{\textbg}[2]{{%
  \setlength{\fboxsep}{1pt}%
  \colorbox{#1}{\vphantom{Xg}#2}%
}}

\definecolor{significant}{HTML}{80B1D3}
\definecolor{notsignificant}{HTML}{FB8072}

\definecolor{likertSD}{HTML}{ca0020}
\definecolor{likertD}{HTML}{f4a582}
\definecolor{likertN}{HTML}{f7f7f7}
\definecolor{likertA}{HTML}{92c5de}
\definecolor{likertSA}{HTML}{0571b0}
\newcommand{\likertLegend}{%
  \textbg{likertSD}{{\color{white}strongly disagree}},
  \textbg{likertD}{disagree},
  \textbg{likertN}{neutral},
  \textbg{likertA}{agree},
  \textbg{likertSA}{{\color{white}strongly agree}}%
}

\newcommand{\subhead}[1]{
  \smallbreak
  \noindent
  \textbf{#1}
}

\newcommand{\subjectiveFast}{the animation was not too fast}
\newcommand{\subjectiveEasy}{the paths of individual points\,/\,clusters are easy to follow}
\newcommand{\subjectiveEasyP}{the paths of individual points are easy to follow}

\newcommand{\subjectiveFrequent}{I would like to use this animation frequently}

\let\mainMessage\subhead

\newdimen\figwidth

}

\begin{document}

\firstsection{Introduction}

\maketitle

Scatter plots are a useful visualization for a pair of numerical dimensions~\cite{Cleveland1988,Utts2005}.
But data often has more than two attributes.
For efficient exploration and analysis in such a case, it is helpful to visualize the same data samples in multiple views that show different dimensions~\cite{nguyen2020evaluation}.

Up to a relatively low number of dimensions, it is possible to visualize the data with combinations of full-sized scatter plots~\cite{nguyen2020evaluation} or with a \acs{SPLOM} for a focus+context~\cite{baudisch2001focus} approach.
However, as the number of data attributes increases, more and smaller plots are necessary.
At the same time, viewers experience a higher cognitive load to link the separately shown data points from many plots in a single coherent mental map.
The \emph{grand tour} tool by Asimov~\cite{asimov1985grand} introduces a sequence of 2D projections to provide viewers with a dense overview of the original data.
Computational methods for \acs{DR}~\cite{jolliffe2002principal, mcinnes2018umap} allow showing data sets with arbitrary dimensionality in a 2D scatter plot.
The resulting \emph{artificial} display dimensions can summarize a data set but are challenging to interpret and trace back to the original data attributes.

We argue that these new artificial dimensions could be treated like existing data dimensions.
The problem for the analysis of low-dimensional data could then be considered as equivalent to the analysis of data with an arbitrary number of dimensions.
Effectively switching between scatter plots of original and artificial dimensions while maintaining a consistent mental map might benefit data analysts and support visual analysis for explainable artificial intelligence (XAI).
Therefore, we advocate the use of animation to show correspondence between scatter plots.

Analysts need to be able to trace data points between the different views---presented sequentially or simultaneously.
There are multiple possible solutions to the issue of perceiving correspondence between dots in different plots.
For instance, if the data set contains only a few samples, the plots could show them with varying shapes or colors without requiring explicit user input.
However, typical humans are not able to remember all intricacies that would be necessary to distinguish dozens of samples~\cite{Miller1956}.

Animation is a well-known technique for users to solve correspondence tasks~\cite{chevalier2014staggering} and supports data sets with hundreds of samples.
As a benefit, other visual variables, e.g., color and shape, remain available to encode additional and better-fitting data attributes.
Various animations for transitions between different scatter plot views were suggested in prior work~\cite{elmqvist2008rolling,du2015trajectory,kim2019designing}, including the \emph{grand tour}~\cite{asimov1985grand}.
We want to explore the design space of animations to identify fitting animation types for the generic tasks of tracing points and cluster interactions between pairs of 2D scatter plots.
We aim for ecological validity beyond tasks with only a few simultaneously traceable data points.
Thus, this paper compares multiple spline-based and 3D-rotation-based transitions in a study with hundreds of data points.

Our main contribution is a preregistered crowdsourced user study with 170 participants.
We designed it to quantitatively evaluate and compare the effectiveness of different animation techniques for accurately tracking individual data points and cluster interactions with a focus on ecological validity.
We analyze the results and report the findings.
Finally, we publish the recorded study data, the source code of the animation framework for transitions between 2D scatter plots (as a plugin for \emph{D3.js}~\cite{bostock2011d3}), and an interactive demo of the study \cite{darusAnim,brandt2023d3,rodrigues2023interactive}.

\section{Related Work}
\label{sec:related-work}

The study behind this paper relates to previous work in multiple respects.
First and foremost, we will discuss animation techniques in general, and for scatter plots in particular.
Then, we will go into studies with animated visualizations as stimuli.
Finally, we will touch on the topic of eye movement in the form of smooth pursuit because participants will be trying to follow moving points and clusters in this study.

\subsection{Animated Visualization in General}

Animation is not a new concept, and there have been multiple publications on the topic.
Moving visualizations are often used to show temporal changes or trends in the underlying data~\cite{Tekusova2007,Robertson2008,Kriglstein2012}.
Gapminder~\cite{gapminder} is one such example, where the movement of dots in a scatter plot encodes the changes of country-level metrics over time.
There are also static visualizations for changes over time that can be used effectively~\cite{Robertson2008,perin2018assessing}.
However, our work does not focus on mapping time to a dedicated third dimension or to the movement of data points.
With this paper, we want to evaluate the use of animation purely for correspondence across the transition between two scatter plots of different attributes from the same data set.

Storytelling is another field of research with multiple publications~\cite{drucker2015sanddance,thompson2021dataanimator,cao2023dataparticles} about the use of animation.
We target the interactive visual exploration and analysis of data sets.
Hence, we will not evaluate the suitability of animation for storytelling but only for tracking individual data points and interactions between clusters.
Animation has also been used to transition between different types of visualization~\cite{drucker2015sanddance,tominski2021flexible}, show how data is aggregated~\cite{kim2019designing}, or display datamations~\cite{guo2023datamator}.
Again, these use cases do not fit our overarching goal behind this paper.

Animated visualization is suited for the analysis of moving sources of data.
This is especially noticeable in the context of sports, where situations from a game can be analyzed post hoc to learn about strategies, strengths, and weaknesses~\cite{footovision}.
However, even in such a case, the animation is tightly linked to the dimension of time.
Yee et al.~\cite{yee2001animated} used animation to support the interactive exploration of graphs with a radial tree layout by linearly interpolating polar coordinates of the nodes while constraining the ordering and orientation of nodes.
Their spiral-shaped animation paths are a good match for radial graph layouts but are not intuitively linked to our targeted use of scatter plots.
  
\subsection{Animations for Scatter Plot Visualizations}
\label{sec:rw:animations-scatter-plot-isualizations}

Thompson et al.~\cite{thompson2021dataanimator} used animation for the correspondence of data points between different plots.
In their work, however, all visible data remained constant, and only the plot type changed.
Our motivation is more related to work like the \emph{grand tour} by Asimov~\cite{asimov1985grand}.
His work produces an animation through a sequence of projections onto 2D planes and is supposed to show the shape of data by taking into account all data dimensions.
As Asimov stated, the \enquote{movies} with the projected data are difficult to understand, even with only four source dimensions, and require extensive experience with the visualization.
Our evaluation targets the correspondence between pairs of the original or projected data dimensions.
Therefore, we only animate between two scatter plots without generating \enquote{movies} that show all data dimensions.
We argue that short transitions and longer periods for observation of a single scatter plot are closer to real-world applications of visual~analytics.

The method by Elmqvist et al.~\cite{elmqvist2008rolling} is a better fit for the goal of our evaluation.
It starts with a regular 2D scatter plot and temporarily extends it by adding a third dimension for depth.
A rotation around an unchanged axis, e.g., height, results in role reversal between the width and depth axes (see the generic concept on the left side of \cref{fig:teaser}).
Afterward, the new depth axis is removed, and the result is simply the exchange of one data dimension for another.
The proposed concept of \enquote{rolling the dice} goes even further and creates sequences of multiple rotations~\cite{elmqvist2008rolling}.
It provides users with controls to navigate the space of all 2D scatter plots within a matrix and preserves brushed data points between views.
Sanftmann and Weiskopf~\cite{sanftmann20123d} proposed a related analysis tool with a technique that is primarily 3D.
It provides a 3D \ac{SPLOM} for navigation as well as a 3D scatter plot for details and only reverts to 2D when showing thumbnails on the surface of the \ac{SPLOM}.
Furthermore, they presented a technique for smooth animations while changing one or two dimensions simultaneously.
The resulting projection leads to consistent motion of dots.
While we implemented 3D rotations of scatter plots, our focus was not on a full-fledged visual analytics tool.

In contrast to the previous approaches, Heer et al.~\cite{Heer2007} identified effective design guidelines for creating transitions between different types of statistical charts, e.g., scatter plots and bar charts.
Kim et al.~\cite{kim2019designing} presented staged transitions between visualization types to depict aggregation.
Both works tackle other problems related to animated plots.
Our study only compares and evaluates which scatter plot transition is most effective for the user.

\subsection{User Studies on Animated Visualizations}
\label{sec:rw:user-studies-animated-visualizations}
  
There is prior work on the evaluation of animated visualizations.
Interaction with, and cognitive effects of, animated visualization are the focus of two qualitative studies by Nakakoji et al.~\cite{Nakakoji2001}.
While they reported animation as beneficial, they also identified core challenges for designing interactive and animated visualization.
Yao et al.~\cite{yao2022visualization} presented a research agenda for visualization in motion and analyzed how well viewers could read data from moving charts.
However, their focus lies on \enquote{entire visualizations moving, rather than data [points] [...] or individual parts of [the] visualization [having] different [...] directions, trajectories, or speeds.}

Animation, i.e., time-varying graphical representation, is often used for visualizing time-oriented data~\cite{Kriglstein2012}.
Financial data, for example, typically appears as time-series and can be animated for analysis.
Tekušová and Kohlhammer~\cite{Tekusova2007} evaluated their system with a usability study that included novices and experts from the financial industry.
Their approach focused on the user experience and system design.
Work by Robertson et al.~\cite{Robertson2008} contains a quantitative evaluation of different methods for visualization of trends in data:
traces, small multiples, and animation.
The results include qualitative feedback and a comparison between the techniques with respect to how accurately study participants can extract information from the underlying data.
Animation is reported to perform worse but provides more enjoyment to the users.
A newer quantitative study by Brehmer et al.~\cite{brehmer2020comparative} also compares animation against small multiples of scatter plots.
Its results confirm that viewers of small multiples complete tasks objectively faster but with subjectively lower confidence.
Our study also uses objective measures for task accuracy.
In addition, we record subjective user ratings to adjust the perceived overall speed and task difficulty.

There are parallels between our work and previous studies with regard to visual primitives and animation techniques.
The study by Chevalier et al.~\cite{chevalier2014staggering} targets the viewer's ability to follow up to three of 30 animated dots.
It compares the condition of simultaneous movement against various movements with temporal offset, i.e., \emph{staggered} animation.
The authors controlled for task difficulty by proxy of various metrics that correlated with task accuracy.
The study results show that \enquote{staggering has a negligible, or even negative, impact on multiple object tracking performance.}
Another study, by Du et al.~\cite{du2015trajectory}, compares straight paths against bundled paths for animated dots.
The results indicate that bundling is of benefit when working with more data points or when there is a higher level of occlusion between dots.
Our study is similar because it identifies groups of data points and uses time offset and trajectory bundling to animate dots between scatter plot views.
However, we examine more types of animations and use a real-world data set with more samples to get results that are more representative of visual analytics tasks.
We strive for ecological validity beyond relatively small laboratory setups.

Our work is related to that of Dragicevic et al.~\cite{dragicevic2011temproal}. They studied object tracking performance in scatter plots for different temporal distortions, e.g., constant speed transitions, slow-in/slow-out, etc.
In our study, we do not vary the temporal distortion.
We are more interested in exploring differences in the animation paths, but also include a temporal offset technique without changing the distortion type.
Wang et al.\cite{wang2018vector} conducted a study with an animation design for transitions between scatter plots.
Their technique is based on vector fields to create smooth, non-linear transitions for clustered data with support for manual sketching.
Wang et al.\ used computational metrics and conducted a user study to compare their approach to trajectory bundling and straight line transitions.
We want our study to yield ecologically valid results without relying on theoretic metrics and we want to include rotation transitions, which might be more intuitive for users.
The latter might be highly relevant for widespread adoption in the field.

Our goal is not to provide a new analysis method for information gain but to help humans make better use of the already existing and intuitive scatter plots.
We analyze the use of animation for showing correspondence between data points in two static visualizations.
Therefore, we restrict the evaluation to the traceability of visual elements and do not include tasks that require interpretation of the underlying data.

\subsection{Studies on Smooth Pursuit Eye Movements}
\label{sec:rw:studies-smooth-pursuit-eye-movements}
Our study requires participants to follow moving points and clusters.
As some visual targets have constant movement, we expect the onset of smooth pursuit eye movements~\cite{robinson1965mechanics} for parts of the transitions.
Smooth pursuit maintains a static image of a moving target on the fovea.
This allows the environment of the target to change, which might lead to motion blur.
Previous eye-tracking studies showed that the movement direction of targets has an impact on various metrics related to smooth pursuit~\cite{rottach1996comparison,Mcilreavy2019}.
Our study is not primarily focused on the analysis of the human ability to control such smooth pursuit eye movements, and in \cref{sec:hypotheses}, we describe how we try to avoid large influences from smooth pursuit.
While we do not expect it, there might be an effect of animation direction on the performance of tracking a visual target.

\section{Design Space}

This section describes the design space for modeling animated transitions between different scatter plots of the same data set.

\subsection{Dimensions and Views}

\def\P{P}
\def\PS{P_{s}}
\def\PE{P_{e}}

For a data set with a total of $D \in \mathbb{N}$ attributes, we define the combination of two dimensions $x$ and $y$ as $D_x \times D_y$.
A view change from $D_1 \times D_2$ to $D_3 \times D_4$ does not introduce or remove data points.
This change only alters their position in the scatter plot from $\PS$ to $\PE$ with $\P \in$ display dimensions $D_x \times D_y$.
Therefore, the data points can be animated on a path between $\PS$ and $\PE$ when transitioning to another combination of data dimensions in the scatter plot.
This technique avoids sudden changes in position and helps preserve the viewer's mental map.
Since scatter plots only visualize two data dimensions, there are two types of plot changes from the user's perspective:

\subhead{1D Transitions}only exchange one of the two plot dimensions for a new data dimension.

\subhead{2D Transitions}exchange both dimensions of the plot.

\subsection{Transitions}
\label{sec:transitions}

This subsection describes the types of transitions we evaluated in our quantitative user study.
Note that there are many hypothetical possibilities to explore the design space and vary the animations, but the practical execution and analysis of a user study become increasingly complex when the number of variations increases.
Therefore, we limit our study to two main types of dot movement with three subtypes each: (1)~\emph{spline-based transitions} and (2)~\emph{rotation-based transitions}~\cite{elmqvist2008rolling}.

\begin{figure}
    \vspace*{1ex}
    \centering
    \def\width{0.3\linewidth}
    \begin{subfigure}{\width}
        \centering
        \includegraphics[width=\textwidth]{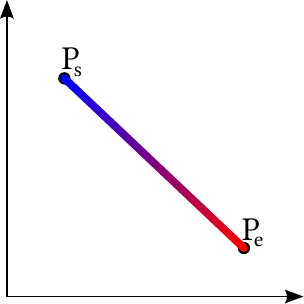}
        \caption{straight}
        \label{fig:paths:straight}
    \end{subfigure}
    \hfill
    \begin{subfigure}{\width}
        \centering
        \includegraphics[width=\textwidth]{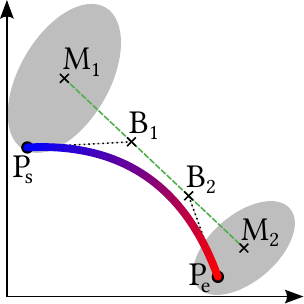}
        \caption{bundled}
        \label{fig:paths:bundled}
    \end{subfigure}
    \hfill
    \begin{subfigure}{\width}
        \centering
        \includegraphics[width=\textwidth]{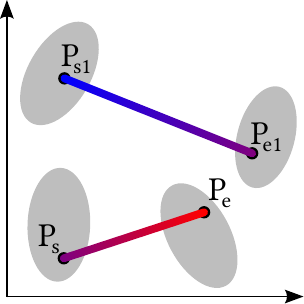}
        \caption{time offset}
        \label{fig:paths:time}
    \end{subfigure}%
    \vspace*{-1.5ex}
    \caption{
        Examples of spline-based animation paths for point movement in transitions between scatter plots.
        Gray ellipses symbolize the area covered by clusters.
        Time is encoded as path color (start~\timeScale~end).
    }
    \label{fig:paths}
\end{figure}

\begin{figure}
  \vspace*{1.5ex}
  \centering
  \includegraphics[width=.27\linewidth]{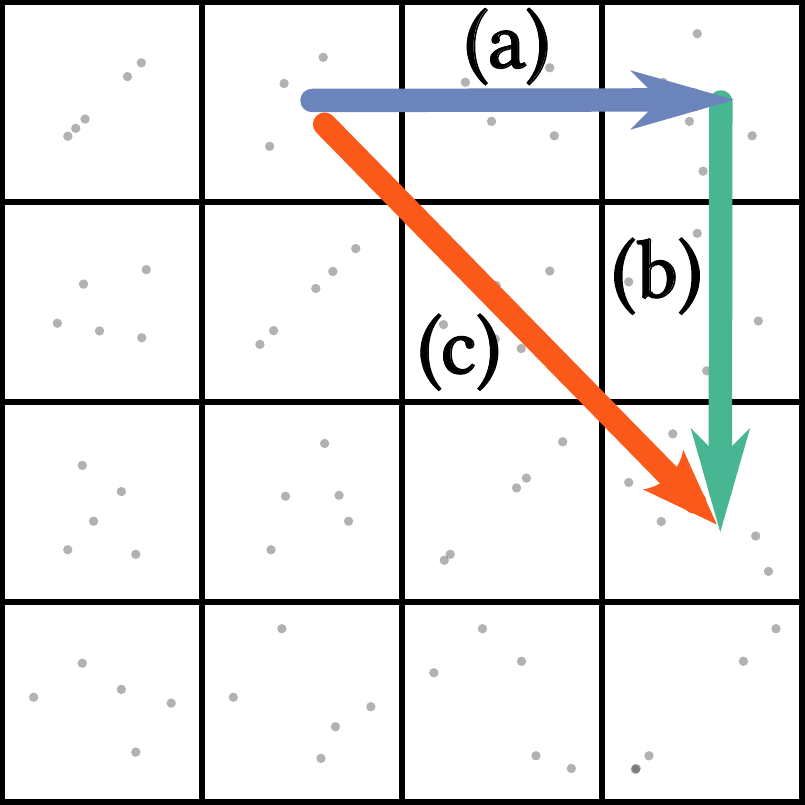}
  \vspace*{-0.5ex}
  \caption{
    1D and 2D transitions in a SPLOM.
    Each square represents a cell of the SPLOM.
    Arrow (a) depicts a horizontal 1D transition, (b) a vertical 1D transition, and (c) a 2D transition.
  }
  \label{fig:hor-ver-both}
\end{figure}

\subsubsection*{Spline-based Transitions}

All spline-based transitions work independently of the number of exchanged dimensions, i.e., whether it is a 1D or 2D transition, since they only require a start point $\PS$ and end point $\PE$.

\subhead{\straightIcon~Straight lines (\straightAbbrev)}are the most basic type of movement and a special case of a spline-based approach.
The path between $\PS$ and $\PE$ is determined by simple linear interpolation, resulting in a straight line (see \cref{fig:paths:straight}).
In the case of 1D transitions, the points move parallel to the changed dimension.

\def\CS{C_s}%
\def\CE{C_e}%
\def\MS{M_s}%
\def\ME{M_e}%
\subhead{\bundledIcon~Bundled lines (\bundledAbbrev)}add complexity over the straight approach by expanding the transition design into the spatial dimension (see \cref{fig:paths:bundled}).
They use clustering and control points to create splines.
In the first step, the DBSCAN algorithm determines a set of clusters $\CS$ and $\CE$ in the start and end pair of normalized data dimensions.
Secondly, we group all data points that belong to the same pair of clusters $C_1 \times C_2 \in \CS \times \CE$.
We calculate the mean position $M_1$ and $M_2$ of all points in $C_1$ and $C_2$.
For each group, we select a set of control points $B$ on the straight line between $M_1$ and $M_2$.
Finally, we compose the splines for each data point's path in the transition by combining the individual start and end positions with the common control points of each group.

\subhead{\timeoffsetIcon~Time offset (\timeoffsetAbbrev)}expands the transition design into the temporal dimension (see \cref{fig:paths:time}).
This method combines the technique from \straight\ and information about cluster pairs $C_1 \times C_2$ from \bundled.
However, each pair of clusters is animated one after the other at a different time offset, such that each cluster is animated separately.
Contrary to prior work~\cite{dragicevic2011temproal}, we only apply an offset and do not vary the temporal distortion between individual dot movements.
While research into the adaptation of the distortion would be of interest, it would also increase the study size with additional variables.

\subsubsection*{Rotation-based Transitions}

\begin{figure}
\vspace*{1ex}
    \centering
    \includegraphics[width=\linewidth]{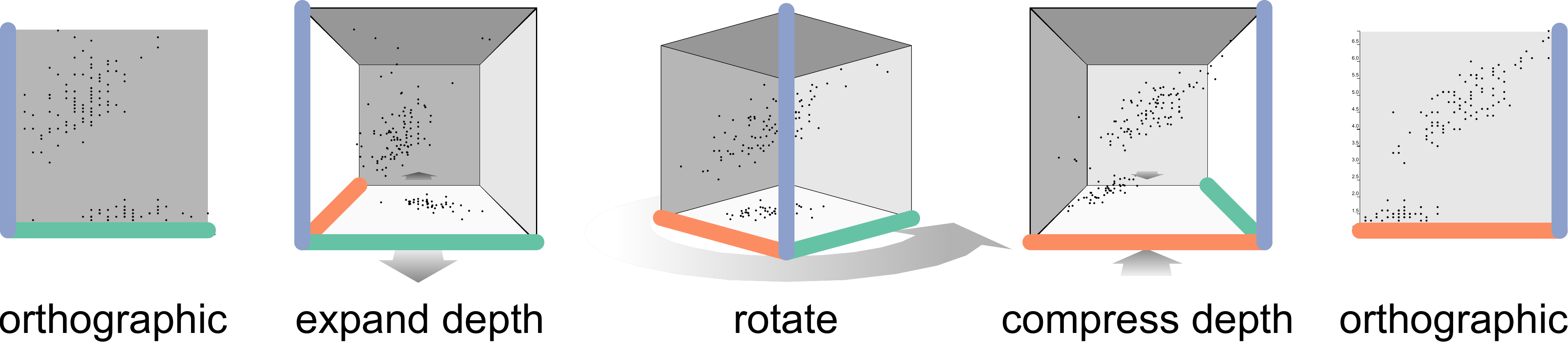}
    \vspace*{-1.5ex}
    \caption{
        Staged rotation (\staged) with a 3D cube happens in three sequential stages (left to right).
        The scatter plots at the beginning and end of the transition share one data dimension.
        The thick colored lines represent data dimensions during the transition.
    }
    \label{fig:rolling-the-dice}
\end{figure}

\begin{figure}
    \vspace*{1.5ex}
    \centering
    \includegraphics[width=\linewidth]{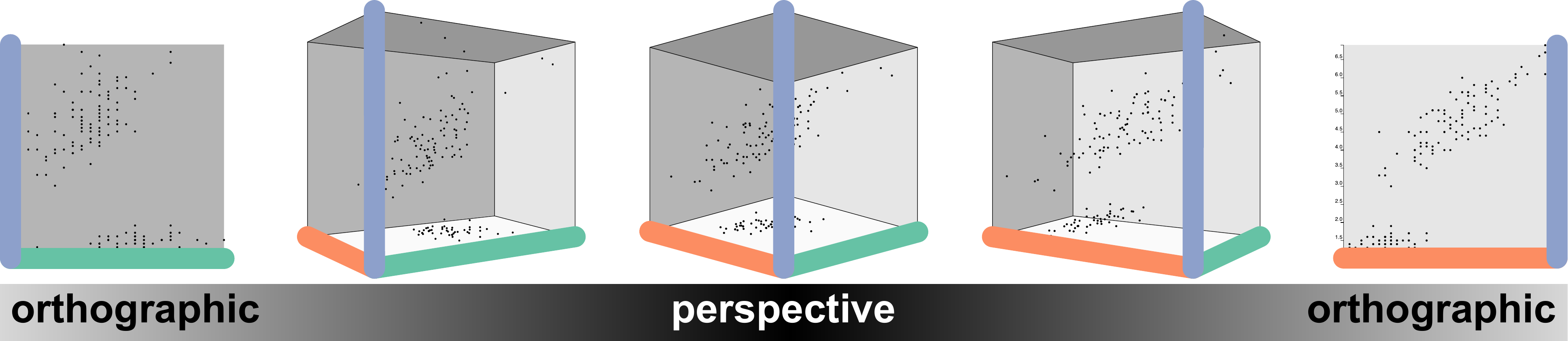}
    \vspace*{-1.5ex}
    \caption{
        Perspective rotation (\perspective) has no dedicated stages (left to right).
        It gradually changes from an orthographic to a perspective camera and back again while simultaneously rotating the cube.
        The thick, colored lines represent data dimensions during the transition.
    }
    \label{fig:perspective-rotation}
\end{figure}

All rotation-based transitions use a 3D cube.
To perform a 1D transition, we encode the new data dimension as depth and then rotate the cube around the unchanged dimension (see \cref{fig:teaser}, left).
We realize 2D transitions through a sequence of 1D transitions: vertical first, then horizontal; or horizontal first, then vertical.
In the context of \acp{SPLOM}, this method corresponds to moving along columns and rows to arrive at the desired view (see \cref{fig:hor-ver-both}, arrows (a) and (b)).

\subhead{\stagedIcon~Staged rotation (\stagedAbbrev)}is our implementation of the method by Elmqvist et al.~\cite{elmqvist2008rolling}.
This prior work is at the core of our selection of rotation transitions and motivates our comparative study.
\Cref{fig:rolling-the-dice} shows the individual steps of the transition technique.
As preparation, we encode the new data dimension as depth ($z$-axis).
However, this depth is not perceivable because we use an orthographic camera.
In the first stage of animation, we smoothly change the values in the camera's projection matrix to form a perspective camera.
This turns the depth information visible and reveals the new data dimension as depth dimension ($z$-axis) of the scatter plot.
Rotation of the 3D scatter plot within the graphics scene is the second stage.
The third and final stage collapses the exchanged $z$-axis by gradually returning to an orthographic camera matrix, while the data points remain static within the graphics scene.

\subhead{\perspectiveIcon~Perspective rotation (\perspectiveAbbrev)}works similarly to \staged\ rotation but without stages.
The extension and collapse of the depth axis and the rotation of the 3D scatter plot take place simultaneously.
\Cref{fig:perspective-rotation} shows the concept.

\subhead{\orthographicIcon~Orthographic rotation (\orthographicAbbrev)}never uses a perspective camera.
We only rotate a 3D scatter plot in the graphics scene to exchange the depth axes.
However, the $z$-axis component of the data points is imperceptible with the orthographic camera.
Therefore, this technique is a simplified version of the original rotation approach that lacks an explicit display of depth information and has no need for individual stages.

The source code published in the supplemental material~\cite{darusAnim} contains details about the used cameras and their parameters, allowing for the construction of the transformation matrices used in computer graphics.

\subsection{Non-evaluated Alternatives}

Our \emph{D3.js}~\cite{bostock2011d3} plug-in provides a framework for further animations.
Splines could theoretically animate points between more than two scatter plots simultaneously.
Instead of using cluster-related control points, we could use the positions of data points in intermediate plots as control points.
This would even allow for animations similar to the \emph{grand tour}~\cite{asimov1985grand}.
However, the longer the animation, the more opportunities to lose track of data points or clusters of interest.
The latter is especially true as the clusters change between the individual subspaces.
We want to evaluate the animation techniques with a fair and practically relevant setup that allows analysts to get an overview of the data and extract finer details.
Our selection of transitions matches the tasks of tracing individual points and clusters.
While the work of Elmqvist et al. inspires the rotation animations~\cite{elmqvist2008rolling}, we specifically designed the \bundled\ and \timeoffset\ techniques to potentially help the viewer to trace clusters.
We argue against the use of both bundling and time-offsets with the rotation animations because they might interfere with the perception of the rotation effect and confuse viewers.
\section{Hypotheses}
\label{sec:hypotheses}

Now that we have a list of six animated transitions, we want to evaluate them comparatively.
Our assessment focuses on accuracy and disregards completion time because all animations have a constant non-user-adjustable runtime.
Analysts extract information from data visualizations with various methods.
While animation could encode a data dimension by itself, e.g., by showing changes over time, the selected transitions are only meant to help preserve the analyst's mental model of the data.
We state the following four hypotheses to explore this notion more precisely.

We aim to evaluate the chosen animations concerning how well a single data point can be traced from one plot to the next.
Thus, we pose the following hypothesis to evaluate the tracing accuracy.

\subhead{Hypothesis \hypPoint} (P for point task): \emph{There is an animation-dependent difference in task accuracy for point tracing.}

\smallskip

It is relatively easy to find clusters in the 2D subspace of a single scatter plot.
But it is nontrivial to trace how the clusters evolve between two scatter plots with different axes.
Other approaches, such as cluster-flow parallel coordinates~\cite{rodrigues2020cluster}, can show how the clusters interact between 2D subspaces.
We suspect that animation can serve the same purpose while avoiding the need to learn a new visualization technique.
Therefore, our work aims to evaluate how well interactions between clusters of different scatter plots can be traced using the selected animations.

\subhead{Hypothesis \hypCluster} (C for cluster task): \emph{There is an animation-dependent difference in cluster task accuracy.}

\smallskip

Previous work suggests a difference in the ability of humans to follow objects vertically vs.\ horizontally through eye movement~\cite{rottach1996comparison,Mcilreavy2019,Purves2001} (see also \cref{sec:rw:studies-smooth-pursuit-eye-movements}).
In practice, analysts need to be able to trace points in both directions.
Therefore, we use square scatter plots that are small enough to trace data points without large eye movements.
With these assumptions, we expect a negligible influence on task accuracy from vertical vs.\ horizontal dot movement.
However, statistical tests can only detect a difference.
Hence, we will perform tests for a difference and interpret a lack thereof as an indication of a negligible effect.

\subhead{Hypothesis \hypDirection} (D for direction): \emph{There is no direction-dependent difference in task accuracy.}

\smallskip

In addition to the purely objective task accuracy with points and clusters, we also want to compare the animation techniques concerning subjective user ratings.
We aim to evaluate whether the user sees animated scatter plot transitions as valuable.
We do this in a similar fashion to Robertson et al.~\cite{Robertson2008}.
However, they explored whether users found animation generally beneficial.

\subhead{Hypothesis \hypRating} (R for rating): \emph{There is an animation-dependent difference in subjective user rating.}

\section{Methods}

Our main goal is to comparatively evaluate the different animation techniques for transitions between 2D scatter plots of a multivariate data set.
To this end, we designed and preregistered a study \cite{rodrigues2022evaluation}.
We documented the hypotheses, statistical tests, and sample sizes in the preregistration before collecting and analyzing any data.
In the following, we describe the setup of our study in detail.
The supplemental material~\cite{darusAnim} contains select video recordings and an interactive demonstration of all study tasks.

\subsection{Tasks}

\subhead{Point Task:}
As mentioned in \cref{sec:hypotheses}, we expect the animation to have an impact on how well users can trace a point during the transition between two plots.
To check \hypPoint, we show a scatter plot and highlight a single point.
The participant should then follow this point along its animation path and mark the position where the point came to rest.
The viewer clicks anywhere within the final scatter plot to indicate the estimated position.
We determine the error by measuring the distance between the actual data point and the selected location.

We use the existing auto-mpg data set modified by Quinlan~\cite{quinlan1993combining} to measure task accuracy under realistic conditions.
It has 398 data points with 8 attributes.
Some attributes are continuous (e.g., weight), while others have only discrete values (e.g., number of cylinders).
The latter data type is not particularly well suited for scatter plots or animation as it may lead to a high degree of overplotting and, subsequently, hinder the identification of individual dots.
However, in practice, data does not always fit the visualization technique, and data dimensions with few discrete values are part of realistic use cases.

\subhead{Cluster Task:}
For this study, we identified three types of possible interactions between data clusters in different scatter plots:
(1) no interaction, the cluster \emph{remains} mostly as it was,
(2) the cluster \emph{splits} into multiple smaller clusters, and
(3) the cluster \emph{merges} with another.
Outside of this study, other interactions are theoretically possible. For instance, a new cluster forms, and an existing one dissolves or spreads.
However, they require users to calculate or intuitively estimate how densely the data points need to be packed concerning their surroundings to form a cluster.
Depending on the individual, this might be highly subjective and yield different results.
Thus, we limit our study to the three described interactions between data clusters.

Finding a real data set with only the presented three cluster interactions is not trivial.
A variation in the total number of visible clusters might influence viewers.
Even if they lose track of the cluster they are trying to follow, a decrease in total clusters might suggest a merge.
In order to reduce confounding effects, we use a generative model~\cite{schulz2016generative} to produce consistent cluster interactions.
The resulting data set contains 600 data points that are distributed into 5 clusters (80 points each) and 200 randomly positioned distractors.
The cluster task has more dots than the data set for the point task.
This corresponds to real-world applications, where the targets for analysis of larger data sets often shift from individual points to clusters.

We implement the interactions by moving half the points from one cluster to another.
Depending on the point of view, this is either a \emph{split} or a \emph{merge}.
The other clusters are not involved in any interaction (\emph{remain}).
We choose specific target clusters for the study task to ensure that all three conditions appear equally often.
During the study, the initial view highlights all cluster data points that should be traced.
When the animation begins, all dots revert to the same look.
Afterward, the participants can select the perceived cluster interaction by clicking one of three buttons.
We use the ratio of correct answers as a measurement for the study.

\subsection{Procedure}
\label{sec:procedure}

As preparation for the study, we randomly select combinations of axes and a data point or cluster to trace.
\label{sec:random-axis-selection}
We compose three combinations of start and end views for transitions.
From these, we create one horizontal and one vertical 1D transition.
Then, we generate three more combinations and use them for 2D transitions.
We replicate them for each start direction of rotation-based animations (horizontal first, then vertical; vertical first, then horizontal).
Spline-based animations, however, have no \enquote{first} axis of movement.
Therefore, we generate three additional combinations.
Each animation is tested with 2$\times$3 1D and 2$\times$3 rotation-based or 6 spline-based 2D transitions to provide the necessary number of trials for the subsequent statistical analysis.

We use the same tests for all animations and participants, albeit in random order.
We test the point task and the cluster task in sequence to avoid frequent switches:
12 point tasks for each animation first, then 12 cluster tasks for each animation.
Since every participant gets to perform every task with every animation, we have a within-subjects study design.

It is important to choose the dots for the point task randomly and re-use the same selection with all animations and participants to minimize unwanted side effects.
In addition, our evaluation is comparative between animation techniques and does not provide absolute values.
Nevertheless, the dot density and overdraw around each data point might have a confounding influence on the task difficulty~\cite{chevalier2014staggering,brehmer2020comparative}, and there might be other still unknown factors.
We are not aware of a metric that reliably yields a complete and robust task difficulty for tracing a data point.
Even if there were a metric that worked on a single data point, how would we combine the many values for an entire scatter plot?
Even if there were a metric that worked on a static scatter plot, how would we select a single value when the metric constantly changes during the animation?
What single value would then represent the potentially time-varying task difficulty?
In a nutshell, we have no method to calculate a complete and reliable metric for difficulty and, therefore, mitigate confounding effects through random selection.
Our study design will only show what animation technique leads to better results.
Further research is necessary to reveal the underlying causes of expected differences, which might include dot overlap.

\begin{figure}
    \vspace*{1ex}
    \centering
    \def\width{0.45\linewidth}
    \begin{subfigure}[t]{\width}
        \includegraphics[width=\linewidth]{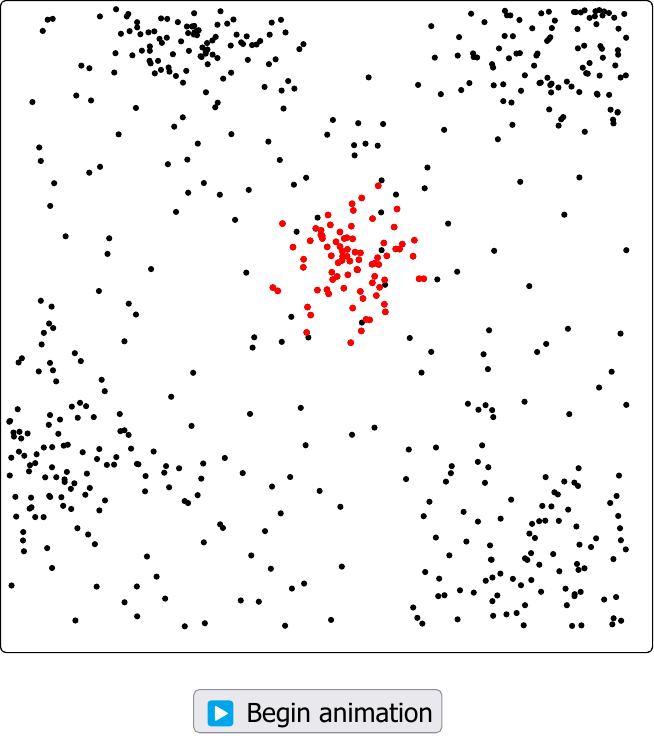}
        \caption{before animation}
        \label{fig:cluster-task:start}
    \end{subfigure}
    \hfill
    \begin{subfigure}[t]{\width}
        \includegraphics[width=\linewidth]{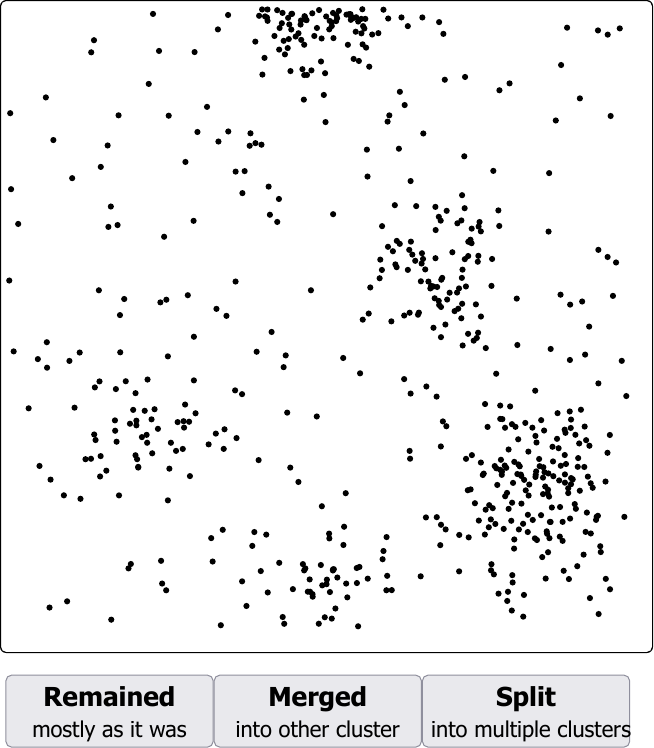}
        \caption{after animation}
        \label{fig:cluster-task:end}
    \end{subfigure}
    \vspace*{-1ex}
    \caption{
        User interface for the cluster task.
        The target cluster is only highlighted in red before the animation.
        After the transition, participants respond whether the cluster \emph{remained} the same, \emph{merged}, or \emph{split}.
    }
    \label{fig:cluster-task}
\end{figure}

During the study, participants are shown a static scatter plot with a highlighted dot or cluster (see \cref{fig:teaser,fig:cluster-task:start}).
They click a button to start the animation.
Afterward, they select a target position and click to submit their answer for the point task.
For cluster-related tasks, the participants single-click one of the buttons representing a cluster interaction (see \cref{fig:cluster-task:end}) to indicate what they have perceived.

We append a short questionnaire to each block of tasks for an animation.
There, we ask for subjective feedback on these statements:
(1) \emakefirstuc{\subjectiveFast}.
(2) \emakefirstuc{\subjectiveEasy}.
(3) \emakefirstuc{\subjectiveFrequent}.
Participants respond on a 5-point Likert scale:
strongly disagree (--2), disagree (--1), neutral (0), agree (+1), strongly agree (+2).
\label{sec:likert-scale}
Generally, we regard the Likert scale as having ordinal values and only use the associated numbers on a rational scale when calculating mean values and standard deviation.

Similar to prior published questionnaires~\cite{hart1988tlx,brooke1996sus}, all statements are phrased in a way that allows the use of a consistent scale.
We acknowledge that there might be effects from participant bias based on the consistently positive phrasing.
However, we argue that the benefits of such a wording in the circumstances of our study outweigh possible drawbacks.
In a crowdsourced setup, where participants rely on the studies to provide income, time is an important factor and quick answers are inherently incentivized.
A consistent Likert scale reduces the mental load for participants to provide answers, thus reducing erroneous inputs.
Most importantly, our analysis is a comparative evaluation.
We do not provide absolute values, and any systemic bias is accounted for by only providing relative differences.
Absolute values would be difficult to interpret, especially as the midpoint of the Likert scale might be misused for undecided answers~\cite{armstrong1987midpoint}.
Additionally, the recorded answers from the main study show no inherent bias toward agreement or disagreement.

\subsection{Training}

At the beginning of each task type, we introduce how the task should be performed and how the participants can submit their answers.
Before each new animation, we provide four task instances as training.
They are generated in the same way as the regular tasks, but we provide feedback after the participant submits a response, i.e., for the point task, we reveal the final position of the dot that should have been followed, and for the cluster task, we show whether the provided answer was correct or wrong.

\subsection{Animation Speed}
\label{sec:speed}

We assume that the speed of the animations has a large effect on the measured accuracy.
All tested animations have the same base duration.
However, we extend the total duration of \staged\ to 250\,\%.
This allows us to maintain a similar rotation speed while providing extra time for the staged switch to and from the perspective camera.
The \timeoffset\ animation gets 200\,\% of the base runtime to limit the speed of the sequential animation of each cluster pair.

\begin{figure}
    \vspace*{1ex}
    \centering
    \setlength\figwidth{\linewidth}
    \begin{subfigure}{\linewidth}
        \centering
        \includegraphics[width=\figwidth]{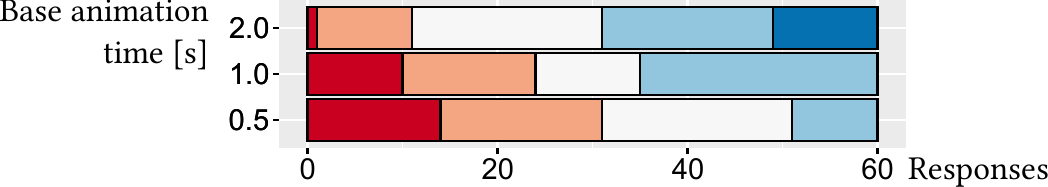}
        \caption{\emakefirstuc{\subjectiveFast}.}
        \label{fig:speed-range}
    \end{subfigure}
    \vspace*{0.5ex}
    \begin{subfigure}{\linewidth}
        \centering
        \includegraphics[width=\figwidth]{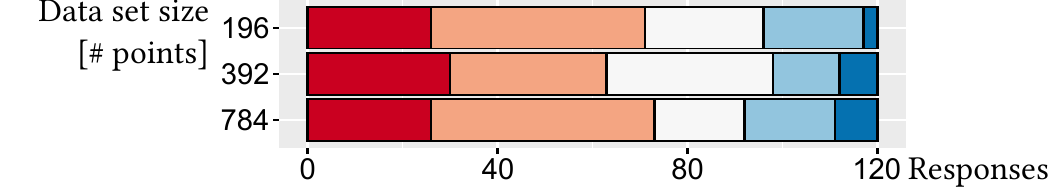}
        \caption{\emakefirstuc{\subjectiveEasyP}.}
        \label{fig:size-range}
    \end{subfigure}
    \vspace*{-3.5ex}
    \caption{
        Results of the pilot studies with different animation speeds \subref{fig:speed-range} and point counts \subref{fig:size-range}.
        Participants rated their agreement with the shown statements on a Likert scale: \likertLegend.
    }
\end{figure}

After the preregistration, we performed a between-subjects pilot study with a similar setup to the main study but varied the movement speeds to get a range of base animation times:
0.5, 1, and 2 seconds.
We had five participants for each condition but did not analyze the actual accuracy data to avoid researcher bias in the main study.
Instead, we compared their subjective feedback regarding the animation speed (\enquote{\subjectiveFast}).
\Cref{fig:speed-range} shows the results.
The variant with 1 second is best suited to find significant differences between the animations because it has similar agreeing and disagreeing scores.
If all participants perceived the speed to be well within their limits, they could trace most data points' paths perfectly.
For too low scores, the selected point positions might be random.

\subsection{Data Set Size}

As mentioned in \cref{sec:procedure}, dot density and overdraw might have a confounding effect and are related to the number of points in the visualized data set.
We argue that an overall rise of dot density has only little effect on cluster tasks but might present a crucial factor for tracing a single point.
Thus, for ecological validity, it is sufficient to examine varying data set sizes for point tracing only.

After the preregistration, we performed a between-subjects pilot study with a similar setup to the point task in the main study but varied the size of the auto-mpg data set used in the plots through random sampling and augmentation:
196 points (50\,\%), the original 392 points (100\,\%), and 786 points (200\,\%).
The resulting data sets were similar in visual appearance and shape, varying mostly in point density.
To get comparable results, we used the same target dots as in the main study and ensured their availability across all conditions.

We had 20 participants for each condition and compared subjective feedback regarding the ease of the task instead of the actual task accuracy to avoid researcher bias.
\Cref{fig:size-range} shows that participant ratings yielded very similar results across all conditions.
While the point tracing task is rated as challenging, the different dot counts have no statistically significant effect on the perceived task difficulty~\cite{darusAnim}.

Previous perceptual studies inferred difficulty from dot density and overdraw around individual points~\cite{chevalier2014staggering,brehmer2020comparative} but had lower overall dot counts than the data sets in our pilot study.
The subjective feedback from our 60 participants suggests that the random selection of target dots is sufficient to control for task difficulty.
We argue that despite low subjective ease, the chosen auto-mpg data set with 392 points is a realistic fit for our main goal of evaluating animated transitions in the context of visual analytics and that the task difficulty might reach saturation as the dot count increases.

\subsection{Power Analysis}
\label{sec:power-analysis}

Prior to preregistration, we performed a power analysis with G*Power 3.1~\cite{faul2009statistical}.
We did not know the shape of the distributions of the measurements a priori.
Therefore, we chose a Wilcoxon signed-rank test for the power analysis of both the point and cluster tasks:
A medium effect size ($dz=0.3$) yielded a sample size of 170 participants.
The Likert scales in the questionnaires require an analysis with $\Chi^2$ tests of independence.
Power analysis for an effect size of $w=0.3$ resulted in a sample size of 207 participants.
We chose to go forward with only 170 participants and accept the altered effect size of $w=0.331$ for the $\Chi^2$ test.
Detailed reports from the power analysis are included in the supplemental material~\cite{darusAnim}.
The assumed statistical tests for the power analysis correspond to the actually performed tests reported in \cref{sec:results}.

\subsection{Participant Recruitment}
\label{sec:participant-recruitment}

We crowdsource participants for our 50-minute study on Amazon Mechanical Turk~\cite{mturk} and Prolific~\cite{prolific}.
They earn approximately 12 EUR per hour.
We ensure at most one submission from each participant on each platform.
Of all participants, 89 are male, 80 female, and one reported \enquote{other or prefer not to say.}
The median age is 30 years, with a standard deviation of 9.19 (recorded in intervals of 10 years).
Our data also includes self-reports on a Likert scale (--2 to 2) about \emph{familiarity} ($Mn=0.31$, $SD=1.21$, $Mdn=1$) and \emph{regular use} of scatter plots ($Mn=-0.78$, $SD=1.05$, $Mdn=-1$).

We employ attention tests to ensure high data quality (see \cref{sec:exclusion-criterion}).
To comply with Prolific's policy for attention checks, we extend each animation-targeted questionnaire with nonsensical items (see supplemental material~\cite{darusAnim} for examples).
When participants arrive at our website, they get a study description and a copy of our data policy.
They may only continue with the study if they consent to the policy.
We ask site visitors not to participate if they do not meet the requirements of (1) normal or corrected to normal vision, (2) no motor impairment, and (3) a computer in a desktop configuration with a pixel-accurate pointing device.
In the second step, the participants adjust the scale of the study interface to ensure comparably sized scatter plots across all~displays.

\section{Results}
\label{sec:results}

In this section, we evaluate various aspects of the collected data according to the hypotheses from \cref{sec:hypotheses}.
We perform various statistical tests and report the results.
Shapiro-Wilk tests showed that the recorded study data is mostly not normally distributed.
Therefore, we mainly use Wilcoxon rank sum tests, as already assumed in the power analysis in \cref{sec:power-analysis}.
Because we inspect the data with regards to different aspects for each hypothesis, we use and report Bonferroni correction.
In the interest of brevity, not all results are included in this section.
We refer the interested reader to the supplemental material~\cite{darusAnim}, which contains the underlying data, all results of the statistical tests, and plots for~analysis.

\subsection{Exclusion Criteria}
\label{sec:exclusion-criterion}

In addition to platform-based rejections (see \cref{sec:participant-recruitment}), we ensure high data quality by including two trivial tasks for each animation and task (24 per participant).
In these, the highlighted elements are well separated from the others and have restricted movement during the animation.
We exclude data from participants if less than half their answers to point checks are within a radius of 0.1 on the normalized scatter plot axes (0 to 1).
Humans are fallible.
Hence, if the regular task results are too accurate, we assume that the study participant tampered with the website or introduced some kind of automation instead of clicking manually.
The threshold for exclusion is a mean offset from the actual dot position of less than one pixel over all point tasks.
We remove submissions with trivial cluster task checks that achieve less than 50\,\% correct responses.
Additionally, we reject submissions that took less time than all animations combined.
We replace rejected submissions by recruiting more participants to ensure precisely 170~samples.

\subsection{Point Task Accuracy}

\begin{figure}
  \centering
  \includegraphics[width=0.9\linewidth]{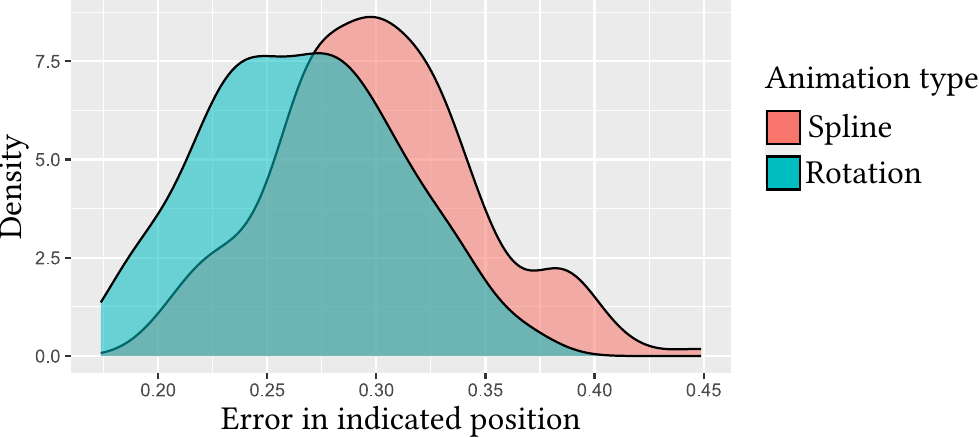}
  \caption{
    Distribution of errors in indicated dot position by animation type.
    Less error indicates better traceability of data points.
  }
  \label{fig:points-by-animation-type}
\end{figure}

We measure the distance between the indicated position and the actual dot within the normalized scatter plot axes $[0,1]$ to determine how well each animation facilitates tracing data points.
We split our investigation of hypothesis \hypPoint\ into multiple aspects.

First, we analyze each animation against every other, using Bonferroni correction.
Shapiro-Wilk normality tests indicate that the distributions for \straight, \timeoffset, and \orthographic\ are not sufficiently close to normal.
Therefore, we use Wilcoxon signed-rank tests to compare individual animations.
\TablerefTabPoints\ in the supplemental material~\cite{darusAnim} show an overview of the results.

We aggregate the study data by animation type to compare spline-based and rotation-based animations (see \cref{fig:points-by-animation-type}).
The aggregation maintains 170 samples because we get one value from the mean positional error across all rotation transitions and one value across all spline transitions for each participant.
The task accuracy with rotation-based animation was better ($Mdn=0.264$) than with splines ($Mdn=0.298$).
An exact Wilcoxon signed-rank test showed that this difference was statistically significant ($p\ll0.0001$, $W=12,536$).
The rank biserial correlation coefficient $rc=0.725$ shows a large effect size.

\subsection{Cluster Task Accuracy}

We use the percentage of correctly identified cluster interactions to compare the animations concerning the traceability of clusters.
The overall mean task accuracy was 84.5\,\%.
First, we perform a pairwise analysis of the task accuracy between the six animations.
Wilcoxon signed-rank tests with continuity correction (there were ties) did not show significant results (see \TablerefTabClusters\ in the supplemental material~\cite{darusAnim}).

\begin{figure}
    \centering
    \includegraphics[width=0.95\linewidth]{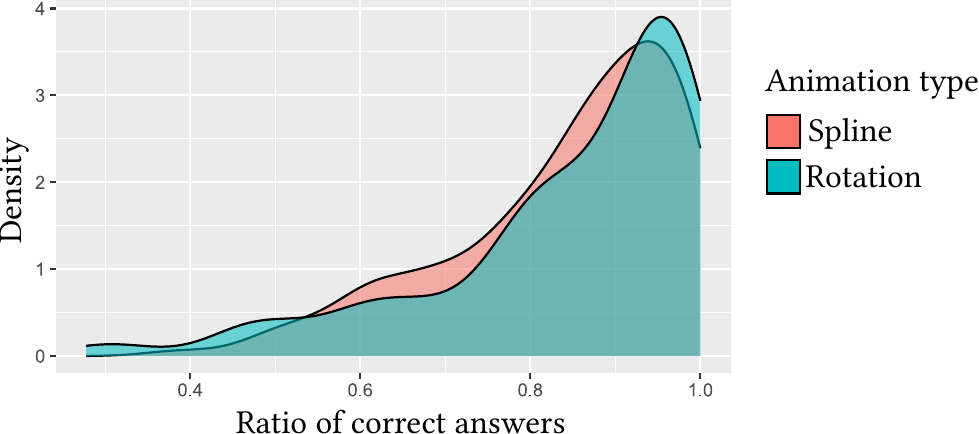}
    \caption{
        Distribution of correctly identified cluster interactions.
        A higher ratio of correct answers indicates better traceability of clusters.
    }
    \label{fig:clusters-by-animation-type}
    \vspace{-3mm}
\end{figure}

In addition to analyzing individual pairs, we also compare the animations grouped by type (see \cref{fig:clusters-by-animation-type}).
Just as with the point task, our averaging method ensures that there are 170 pairs of values (rotation vs.\ spline for each participant).
The task accuracy with rotation-based animation was slightly better ($Mdn=91.6$\,\%) than with splines ($Mdn=87.5$\,\%).
A Wilcoxon signed-rank test with continuity correction did not provide evidence for statistical significance ($p=0.2574$, $W=4,328.5$, $z=-1.12$, $r=-0.086$).
The mean task accuracy of rotation ($Mn=0.847$, $SD=0.158$) and spline ($Mn=0.843$, $SD=0.134$) animations was similar.

\subsection{Animation Direction}

For 1D transitions, the accuracy of the point task was lower with animation in the horizontal direction ($Mdn=0.3994$) than in the vertical direction ($Mdn=0.2682$).
A Wilcoxon signed-rank exact test showed that this difference was statistically significant ($p\ll0.0001$, $W=14,510$, $rc=0.997$).
The pairwise tests between all animations and with Bonferroni correction were also statistically significant (see \TablerefTabDirection\ in the supplemental material~\cite{darusAnim}).

There was no significant difference in the overall accuracy for the cluster task with 1D transitions.
However, for \bundled\ splines, the horizontal accuracy was better than with vertical transitions (pseudo $Mdn=+0.3333$).
A Wilcoxon signed-rank test with continuity correction showed that this difference was statistically significant ($p=0.0047$, $W=1,044$).
The rank biserial correlation coefficient indicated a medium effect size ($rc=0.471$).
For \timeoffset, the horizontal accuracy was, once more, worse than in the vertical direction (pseudo $Mdn=-0.3333$).
A Wilcoxon signed-rank test with continuity correction showed that this difference was statistically significant ($p\ll0.0001$, $W=641$).
The rank biserial correlation coefficient indicated a large effect size ($rc=-0.585$).

\begin{figure}
    \centering
    \def\width{0.85\linewidth}%
    \begin{subfigure}{\width}
        \includegraphics[clip,trim=0 18 0 0,width=\textwidth]{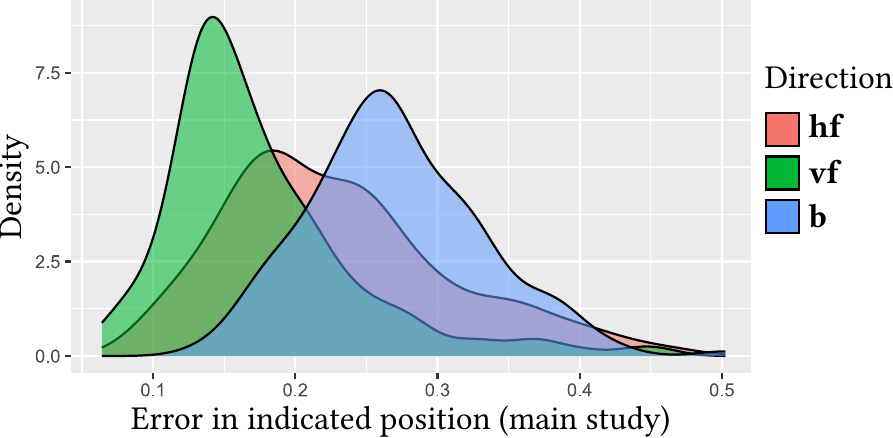}
        \caption{Error in indicated position (main study)}
        \label{fig:2d-direction-points:main}
    \end{subfigure}\\%
    \begin{subfigure}{\width}
        \includegraphics[clip,trim=0 18 0 0,width=\textwidth]{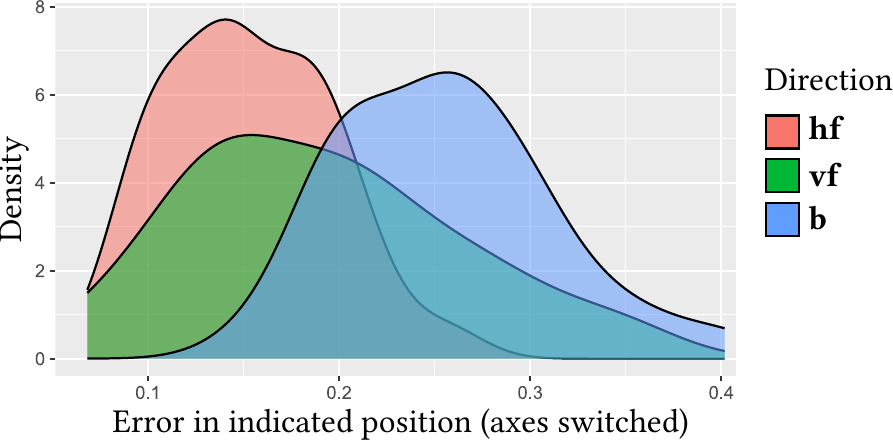}
        \caption{Error in indicated position (axes switched)}
        \label{fig:2d-direction-points:switched}
    \end{subfigure}%
    \vspace{-2mm}
    \caption{
        Distribution of point task accuracy with 2D transitions for each direction.
        Smaller error values are better.
        The upper diagram shows the data from the main study with 170 participants.
        We recorded the data for the lower diagram in a smaller study with switched axes and 30 participants.
    }
    \label{fig:2d-direction-points}
    \vspace{-3mm}
\end{figure}

Directions in 2D transitions differ from the ones available in one dimension.
Rotation-based animations can run \emph{horizontal first} (\hf), then vertical, as well as \emph{vertical first} (\vf), then horizontal.
Spline-based animations always use \emph{both} (\bo) directions simultaneously.
For the point task, \cref{fig:2d-direction-points:main} shows large differences in accuracy.
The error of the point task with 2D transitions was greater in \hf\ direction ($Mdn=0.2190$) than in \vf\ ($Mdn=0.1576$).
A Wilcoxon signed-rank exact test showed that this difference was statistically significant ($p\ll0.0001$, $W=12,380$, $rc=0.703$).
The error of the point task with 2D transitions was less in \hf\ direction ($Mdn=0.2190$) than in \bo\ ($Mdn=0.2646$).
A Wilcoxon signed-rank exact test showed that this difference was statistically significant ($p\ll0.0001$, $W=3,089$, $rc=-0.575$).
There were no significant differences for the cluster task.

We were surprised by the strong effects of the direction of transitions.
They might be linked to the point distributions in the underlying data dimensions.
The random selection (see \cref{sec:random-axis-selection}) could lead to the uneven use of dimensions with few unique values.
Therefore, we repeated the point task with 30 additional participants in a smaller follow-up study.
In this new study, we reused the previous selection of data dimensions but switched their roles as $x$-axis and $y$-axis for the plots.
The change resulted in an automatic mirroring of animation directions.

In the follow-up study, the errors in the 1D point task were lower with horizontal ($Mdn=0.2834$) than with vertical ($Mdn=0.3738$) transitions.
A Wilcoxon signed-rank exact test showed that the difference was statistically significant ($p\ll0.0001$, $W=0$, $rc=-1$).
For the point task with 2D transitions (see \cref{fig:2d-direction-points:switched}), the error in \hf\ direction ($Mdn=0.1458$) was lower than in \vf\ ($Mdn=0.1875$).
A Wilcoxon signed-rank exact test confirmed that the difference was statistically significant ($p=0.01367$, $W=114$, $rc=-0.510$).
Again, animation in \vf\ direction ($Mdn=0.1875$) was less error-prone than in \bo\ direction ($Mdn=0.2529$).
A Wilcoxon signed-rank exact test confirmed that the difference was statistically significant ($p=0.0003$, $W=67$, $rc=-0.712$).
Therefore, the simultaneous animation in both directions performed worst in the main study and the follow-up.

\subsection{Subjective Rating}

\begin{figure}
    \vspace*{1.5ex}
    \centering
    \includegraphics[width=0.97\linewidth]{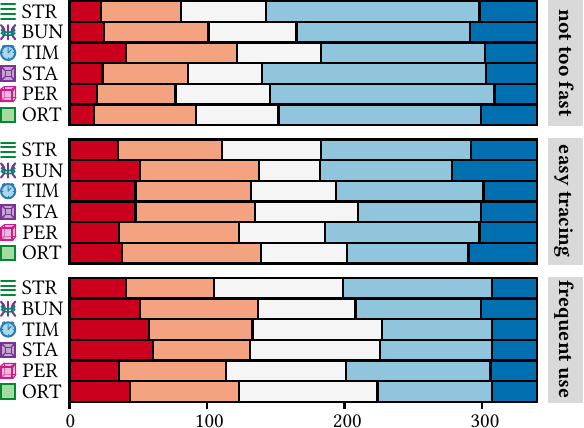}
    \caption{
        Subjective feedback from study participants to the statements from \cref{sec:likert-scale}.
        Responses are given on a Likert scale:\\
        \likertLegend.
    }
    \label{fig:rating-by-animation}
    \vspace{-4mm}
\end{figure}

To compare the animations with regard to subjective feedback from participants, we asked for the study participants' agreement with specific statements using a Likert scale (--2 to 2, see \cref{sec:likert-scale}).
\Cref{fig:rating-by-animation} shows the combined results.
Since we work with ordinal data, we employ $\Chi^2$ tests of independence.
However, to calculate mean values, we use the numeric representation of the Likert scale.

We asked whether they found it easy to trace the points and clusters.
Overall, the participants responded neutrally ($Mn=0.064$, $SD=1.265$).
Transitions were reported to be easier for the point task with \straight\ ($Mn=-0.2$, $SD=1.224$) than with \bundled\ ($Mn=-0.665$, $SD=1.221$).
A $\Chi^2$ test of independence showed that there is a significant relationship between the variables of animation (\straight\ vs. \bundled) and participant rating ($\Chi^2(4, 170)=17.085$, $p=0.0019$).
This is the only significant difference between the rated ease of tracing points or clusters with different animations (see \TablerefTabSubjectivePointsEase\ in the supplemental material~\cite{darusAnim}).

After each point and cluster task, we asked the participants whether they would like to use the animation more frequently.
Overall, the responses were neutral ($Mn=-0.035$, $SD=1.210$).
Participants reported preferring animations with \straight\ ($Mn=-0.171$, $SD=1.167$) over \bundled\ ($Mn=-0.653$, $SD=1.078$) for the point task.
A $\Chi^2$ test of independence showed that there is a significant relationship between the variables of animation (\straight\ vs. \bundled) and participant rating ($\Chi^2(4, 170)=18.052$, $p=0.0012$).
This is the only significant difference between the preferences for animations for tracing points or clusters (see \TablerefTabSubjectivePointsFrequent\ in the supplemental material~\cite{darusAnim}).

In \cref{sec:speed}, we reported on a pilot study that we used to find an animation speed toward which participants were neutral.
The main study also asks to rate the agreement with \enquote{\subjectiveFast} after each point and cluster task.
Overall, the responses were neutral ($Mn=0.313$, $SD=1.137$).
Participants reported with statistical significance that animations with \timeoffset\ were generally too fast for the point task (see \TablerefTabSubjectivePointsFast\ in the supplemental material~\cite{darusAnim}).
There were no significant differences for the cluster task.

\section{Discussion}

With the results from \cref{sec:results}, we can now revisit the hypotheses from~\cref{sec:hypotheses}.

\mainMessage{Significant Differences in Point Task Accuracy}

Hypothesis \hypPoint\ about point task accuracy is confirmed.
It holds for individual pairs of animations but also when we aggregate spline-based and rotation-based transitions.
We use the differences in (pseudo)median errors to create a ranking of animations:
\textbf{\rank{1}~\orthographic, \staged; \rank{2}~\perspective, \straight, \bundled; \rank{3}~\timeoffset}.
Ranks are shared when there is no statistical significance between items.
However, we still list the animations within each rank according to their pairwise difference in (pseudo)medians.
The aggregated results also indicate that the transition type affects the error rate:
\textbf{\rank{1}~rotations; \rank{2}~splines}.
We are surprised that the more complex rotations perform better than the simple straight-line animations---even in 1D.
It might be interesting to investigate possible causes in further studies.

Bonferroni correction is very conservative with regard to type I errors and quickly decreases the threshold for statistical significance.
Without this correction, the difference in task accuracy between \bundled\ vs. \perspective\ and \timeoffset\ (see \TablerefTabPointsAlpha\ in the supplemental material~\cite{darusAnim}) would have been marked as significant.
The resulting hypothetical ranking would have been \rank{1}~\orthographic, \staged; \rank{2}~\perspective, \straight; \rank{3}~\bundled; \rank{4}~\timeoffset.
However, without further dedicated studies, we cannot assume these hypothetical results to be significant.

\mainMessage{No Significant Differences in Cluster Task Accuracy}

There is no evidence to confirm hypothesis \hypCluster\ about accuracy differences in the cluster task.
We found no significant results between individual pairs of animations.
Even when we aggregated the data by transition type (rotation vs.\ spline), the results were significant, but the effect size was below small levels ($|r|=0.086<0.1$).
In general, participants identified cluster interactions well (84.5\,\% correct).
Therefore, scatter plots with animated transitions might provide an intuitive alternative to more complex visualization techniques for cluster interactions.

We suspect that we encountered a ceiling effect~\cite{taylor2010ceiling} because---with the same animation speed---clusters are easier to trace than points because the groups of dots might merge to a single perceptual entity during movement.
It might be necessary to adjust the animation timing or investigate further methods to avoid the ceiling effect~\cite{chyung2020evidence} and to find significant differences.

In practice, however, it is improbable that the transition speeds between scatter plots vary case-by-case, depending on the user's task.
The analyst would have to change settings manually before each animation, or the visual analytics software would need to detect the user's task and intent autonomously.
The latter is an active field of research~\cite{gadhave2021predicting,hajiabolhassani2013computational}, but, at the time of writing, we are not aware of robust solutions that would not lead to user irritation due to wrongly deduced animation speeds.
Therefore, we argue that an ecologically valid comparison of individual animations and transition types requires the same speed for all tasks.

\mainMessage{Possibly Significant Differences Depending on Animation Direction}

Following the results from the main study, we have to reject hypothesis \hypDirection.
Nevertheless, we suspect that the differences in task accuracy between animation directions are primarily due to a combination of the data set and a \enquote{randomly bad} selection of the underlying data dimensions for display in the scatter plots.
A small follow-up study with switched $x$-axis and $y$-axis shows an inversion of the results regarding the differences between horizontal and vertical directions in both 1D and 2D transitions.
The effect sizes $rc$ from both studies are roughly the same in the case of 1D view changes (0.997$\approx$1) and both large for two dimensions (0.703 and 0.51).
The difference in 2D transitions might be a real effect, or it might be a result of uncertainty due to the small sample size in the follow-up study.
The task accuracy was consistently worst when altering both axes simultaneously.
Further research is required to determine whether this is caused by an increase in cognitive or perceptual load or is due to the generally inferior results of spline-based transitions for the point task.
The primary goal of the main study is to evaluate the various animations, not directions.
Please note that the analysis of the presented animations is not affected by possible horizontal or vertical effects because we used the same combinations of start and end views for all transitions.

\mainMessage{Participants Prefer Straight Animation Paths}

For hypothesis \hypRating, we asked participants to state their agreement with the given statements.
Regarding individual animations, there were mostly no significant differences between the agreement of participants with \enquote{\subjectiveEasy} and \enquote{\subjectiveFrequent.}
We can confirm \hypRating\ concerning the direct comparison between \straight\ and \bundled\ lines (preference toward the former) for animation in the point task.
The results for both statements---ease and preference---yield a ranking where \bundled\ is probably last.

Hypothetically, without the restrictive Bonferroni correction, \timeoffset\ would share the last rank with \bundled.
We can confirm \hypRating\ in connection with a reportedly short playtime for the \timeoffset\ animation (last rank).
Without Bonferroni correction, we would have a second-to-last rank for \bundled.
We cannot accept these hypothetical significance values with certainty.
However, they give an indication for evaluations of interest for follow-up studies.

\mainMessage{Worse Feedback for Animations That Highlight Clusters}

Both animation techniques that we designed to highlight clusters were ranked worse.
We suspect this might be due to potentially faster dot movements with \timeoffset\ and more overdraw from higher proximity with \bundled.

Note that the objectively measured error in point tracing shows significant differences between animations, but the subjective rating does not.
We suspect that humans might not estimate well how good they are at visually tracking an object when there are multiple distractors.
We did not provide feedback to the participant about the positional error to avoid learning effects---remember that we used the same marked points and data dimensions with all animations for better comparability.
We speculate that there might be a connection to the results of the study by Robertson et al.~\cite{Robertson2008}, where participants liked interacting with animations independently of accuracy.
Further analyses, however, are outside of the scope of this paper.

\mainMessage{Base Animation Time of 1 Second is Sufficient}

The overall results from the main study confirm the findings of the pilot study: we chose the animation time so that, on average, the participants found the dot movements neither too quick nor too slow.
This resulted in a base animation time of 1 second, which was sufficient for participants to effectively track individual points and cluster interactions alike.
We recommend this duration as a sensible default and encourage the use of animation in visual analytics software that provides ways of switching between different scatter plots of the same data set, e.g., via \acp{SPLOM}.
To accommodate a wider range of users and their preferences, animation should not be mandated.

\mainMessage{Recommendation for Orthographic Rotation}

In real-world applications, users are limited by hard tasks, not by trivial ones.
Since participants solved the cluster task mostly correctly with all animations, the point task is decisive for the evaluation of the animated transitions.
We recommend the use of \orthographicIcon~orthographic rotation based on the ranking from the accuracy of tracing points.
\stagedIcon~Staged rotation, implemented to resemble the animation from related work~\cite{elmqvist2008rolling}, shares the first rank but requires a longer playtime due to the three sequential stages.
Our recommendation is in accordance with the subjective user preference, which was only directed against \bundled\ and \timeoffset\ lines.

\section{Limitations}

After conducting the study and analyzing the results, we noticed limitations of the study design.
We regard a common animation speed for the point tracing and the cluster interaction task as being realistic for currently available visual analytics software.
For research purposes, however, it could have been more interesting to investigate independent animation times in order to get more diverse results for the simpler cluster task.
We suspect that a speed setting that yields a ratio of correct answers around 50\,\% could be better suited to find significant results when comparing different animation techniques for cluster interactions.

The results from the main study indicate a large effect from the animation direction.
In fact, it seems much larger than related work would suggest.
Data from the follow-up study did provide evidence against such a large effect but was not able to confirm or reject our initial hypothesis \hypDirection.
Our primary goal was not the comparison of directions but of the animation techniques.
This is reflected in the study design.
To definitively answer \hypDirection, it would be necessary to record samples for both the original combinations of data dimension as well as the $x$-$y$-switched variants.
We opted against such a design to get higher-quality data for the comparison of animation techniques.
We argue that it would have been counter-productive to pursue more goals because the study duration was already 50 minutes.
A study time of 100 minutes (double each task) could present a higher threshold for recruiting crowdworkers and might lead to high degrees of fatigue in participants.
While the presented scatter plots were small, we allowed data points to appear outside the plot bound during the animation.
In practice, not all use cases provide unused space around the scatter plot, e.g., on mobile devices.

\section{Conclusion}

We developed a framework for animated transitions between scatter plots of different dimensions from the same data set.
We implemented various animations and provided an open-source \emph{D3.js} plug-in.
We designed, preregistered, and conducted a crowdsourced user study with 170 participants to evaluate the different transitions.
The results show that rotation-based animation outperforms its spline-based counterpart.
\orthographicIcon~Orthographic and \stagedIcon~staged rotation worked best to allow participants to trace individual data points across transitions.
Note, however, that staging requires more runtime.
Cluster-centric approaches of spline-based animation did not work well for the point task.
All transitions resulted in good and similar accuracy for the cluster interaction task.

Based on these results, we encourage using animated transitions between different views of the same underlying data set.
With only a small impact on time, they provide effective ways to trace individual points and clusters.
As a result, analysts can make more effective use of point-based visualization in scatter plots without using additional visual attributes for correspondence.
Animation seems to be an intuitive approach because participants were able to follow the target elements across changing data dimensions without requiring lengthy training or extensive experience in the field.
More experienced data analysts could benefit from correspondence through animation when transitioning between actual data dimensions and the abstract results from dimensionality reduction.

Future research might identify and quantify effects from the direction of animated transitions.
We required participants of our study to use desktop-like computer setups, but it would be interesting to reevaluate the task accuracy with animations on smaller and moving displays.
Further research could explore transitions that account for more than two scatter plots (supported in our framework through intermediate views and control points for splines).
A follow-up study could avoid the ceiling effect in the cluster task and compare the accuracy between both tasks.
Future work might look into how data correlations can be tracked across animated transitions of scatter plots and how our recommendation for more animation could be applied to other types of visualization.
Similarly to the \emph{grand tour}~\cite{asimov1985grand}, one might also provide overview videos of the data set with sequences of dimension pairs that are optimized for the specific animations.

\acknowledgments{%
  Funded by the Deutsche Forschungsgemeinschaft (DFG, German Research Foundation) -- Project-ID 251654672 -- TRR 161 (Projects A03 and B01).

}

\bibliographystyle{abbrv-doi}

\bibliography{ms}

\begin{thebibliography}{10}

\bibitem{mturk}
{Amazon Mechanical Turk, Inc.}
\newblock {Amazon Mechanical Turk}.
\newblock \url{https://www.mturk.com/}, accessed 2023-09-16.

\bibitem{armstrong1987midpoint}
R.~L. Armstrong.
\newblock The midpoint on a five-point {L}ikert-type scale.
\newblock {\em Perceptual and Motor Skills}, 64(2):359--362, 1987. doi: {{%
10\hspace{.1pt}\discretionary{.}{%
}{.}\hspace{.4pt}2466\discretionary{/}{%
}{/}pms\hspace{.1pt}\discretionary{.}{%
}{.}\hspace{.4pt}1987\hspace{.1pt}\discretionary{.}{%
}{.}\hspace{.4pt}64\hspace{.1pt}\discretionary{.}{%
}{.}\hspace{.4pt}2\hspace{.1pt}\discretionary{.}{%
}{.}\hspace{.4pt}359}}


\bibitem{asimov1985grand}
D.~Asimov.
\newblock The grand tour: A tool for viewing multidimensional data.
\newblock {\em {SIAM} Journal on Scientific and Statistical Computing}, 6(1):128--143, 1985. doi: {{%
10\hspace{.1pt}\discretionary{.}{%
}{.}\hspace{.4pt}1137\discretionary{/}{%
}{/}0906011}}


\bibitem{baudisch2001focus}
P.~Baudisch, N.~Good, and P.~Stewart.
\newblock Focus plus context screens.
\newblock In {\em Proceedings of the {ACM} Symposium on User Interface Software and Technology}. {ACM}, 2001. doi: {{%
10\hspace{.1pt}\discretionary{.}{%
}{.}\hspace{.4pt}1145\discretionary{/}{%
}{/}502348\hspace{.1pt}\discretionary{.}{%
}{.}\hspace{.4pt}502354}}


\bibitem{bostock2011d3}
M.~Bostock, V.~Ogievetsky, and J.~Heer.
\newblock {$D^3$}: Data-driven documents.
\newblock {\em {IEEE} Transactions on Visualization and Computer Graphics}, 17(12):2301--2309, 2011. doi: {{%
10\hspace{.1pt}\discretionary{.}{%
}{.}\hspace{.4pt}1109\discretionary{/}{%
}{/}TVCG\hspace{.1pt}\discretionary{.}{%
}{.}\hspace{.4pt}2011\hspace{.1pt}\discretionary{.}{%
}{.}\hspace{.4pt}185}}


\bibitem{brandt2023d3}
V.~Brandt and N.~Rodrigues.
\newblock D3-plugin for animated scatter plot transitions.
\newblock \url{https://github.com/NilsRodrigues/d3-scattertrans}, accessed 2024-01-04, 2024.

\bibitem{brehmer2020comparative}
M.~Brehmer, B.~Lee, P.~Isenberg, and E.~K. Choe.
\newblock A comparative evaluation of animation and small multiples for trend visualization on mobile phones.
\newblock {\em {IEEE} Transactions on Visualization and Computer Graphics}, 26(1):364--374, 2020. doi: {{%
10\hspace{.1pt}\discretionary{.}{%
}{.}\hspace{.4pt}1109\discretionary{/}{%
}{/}tvcg\hspace{.1pt}\discretionary{.}{%
}{.}\hspace{.4pt}2019\hspace{.1pt}\discretionary{.}{%
}{.}\hspace{.4pt}2934397}}


\bibitem{brooke1996sus}
J.~Brooke.
\newblock {SUS}: {A} `quick and dirty' usability scale.
\newblock In P.~W. Jordan, B.~Thomas, B.~A. Weerdmeester, and I.~L. McClelland, eds., {\em Usability Evaluation in Industry}, pp. 189--194. Taylor \& Francis, London, 1996. doi: {{%
10\hspace{.1pt}\discretionary{.}{%
}{.}\hspace{.4pt}1201\discretionary{/}{%
}{/}9781498710411}}


\bibitem{cao2023dataparticles}
Y.~Cao, J.~L. {E}, Z.~Chen, and H.~Xia.
\newblock {DataParticles}: {B}lock-based and language-oriented authoring of animated unit visualizations.
\newblock In {\em Proceedings of the {CHI} Conference on Human Factors in Computing Systems}. {ACM}, 2023. doi: {{%
10\hspace{.1pt}\discretionary{.}{%
}{.}\hspace{.4pt}1145\discretionary{/}{%
}{/}3544548\hspace{.1pt}\discretionary{.}{%
}{.}\hspace{.4pt}3581472}}


\bibitem{chevalier2014staggering}
F.~Chevalier, P.~Dragicevic, and S.~Franconeri.
\newblock The not-so-staggering effect of staggered animated transitions on visual tracking.
\newblock {\em {IEEE} Transactions on Visualization and Computer Graphics}, 20(12):2241--2250, 2014. doi: {{%
10\hspace{.1pt}\discretionary{.}{%
}{.}\hspace{.4pt}1109\discretionary{/}{%
}{/}tvcg\hspace{.1pt}\discretionary{.}{%
}{.}\hspace{.4pt}2014\hspace{.1pt}\discretionary{.}{%
}{.}\hspace{.4pt}2346424}}


\bibitem{chyung2020evidence}
S.~Y.~Y. Chyung, D.~Hutchinson, and J.~A. Shamsy.
\newblock Evidence-based survey design: Ceiling effects associated with response scales.
\newblock {\em Performance Improvement}, 59(6):6--13, 2020. doi: {{%
10\hspace{.1pt}\discretionary{.}{%
}{.}\hspace{.4pt}1002\discretionary{/}{%
}{/}pfi\hspace{.1pt}\discretionary{.}{%
}{.}\hspace{.4pt}21920}}


\bibitem{Cleveland1988}
W.~S. Cleveland and M.~E. McGill, eds.
\newblock {\em Dynamic Graphics for Statistics}.
\newblock Wadsworth \& Brooks/Cole Advanced Books \& Software, Pacific Grove, 1988.

\bibitem{dragicevic2011temproal}
P.~Dragicevic, A.~Bezerianos, W.~Javed, N.~Elmqvist, and J.-D. Fekete.
\newblock Temporal distortion for animated transitions.
\newblock In {\em Proceedings of the International Conference on Human Factors in Computing Systems}, pp. 2009--2018. {ACM}, 2011. doi: {{%
10\hspace{.1pt}\discretionary{.}{%
}{.}\hspace{.4pt}1145\discretionary{/}{%
}{/}1978942\hspace{.1pt}\discretionary{.}{%
}{.}\hspace{.4pt}1979233}}


\bibitem{drucker2015sanddance}
S.~Drucker and R.~Fernandez.
\newblock A unifying framework for animated and interactive unit visualizations.
\newblock Technical Report MSR-TR-2015-65, Microsoft Research, 2015.

\bibitem{du2015trajectory}
F.~Du, N.~Cao, J.~Zhao, and Y.-R. Lin.
\newblock Trajectory bundling for animated transitions.
\newblock In {\em Proceedings of the {CHI} Conference on Human Factors in Computing Systems}. {ACM}, 2015. doi: {{%
10\hspace{.1pt}\discretionary{.}{%
}{.}\hspace{.4pt}1145\discretionary{/}{%
}{/}2702123\hspace{.1pt}\discretionary{.}{%
}{.}\hspace{.4pt}2702476}}


\bibitem{elmqvist2008rolling}
N.~Elmqvist, P.~Dragicevic, and J.-D. Fekete.
\newblock Rolling the dice: Multidimensional visual exploration using scatterplot matrix navigation.
\newblock {\em {IEEE} Transactions on Visualization and Computer Graphics}, 14(6):1539--1148, 2008. doi: {{%
10\hspace{.1pt}\discretionary{.}{%
}{.}\hspace{.4pt}1109\discretionary{/}{%
}{/}TVCG\hspace{.1pt}\discretionary{.}{%
}{.}\hspace{.4pt}2008\hspace{.1pt}\discretionary{.}{%
}{.}\hspace{.4pt}153}}


\bibitem{faul2009statistical}
F.~Faul, E.~Erdfelder, A.~Buchner, and A.-G. Lang.
\newblock Statistical power analyses using {G*Power} 3.1: Tests for correlation and regression analyses.
\newblock {\em Behavior Research Methods}, 41(4):1149--1160, 2009. doi: {{%
10\hspace{.1pt}\discretionary{.}{%
}{.}\hspace{.4pt}3758\discretionary{/}{%
}{/}brm\hspace{.1pt}\discretionary{.}{%
}{.}\hspace{.4pt}41\hspace{.1pt}\discretionary{.}{%
}{.}\hspace{.4pt}4\hspace{.1pt}\discretionary{.}{%
}{.}\hspace{.4pt}1149}}


\bibitem{footovision}
FOOTOVISION.
\newblock {Footovision: Level up your game}.
\newblock \url{https://www.footovision.com/}, accessed 2023-07-09, 2023.

\bibitem{gadhave2021predicting}
K.~Gadhave, J.~Görtler, Z.~Cutler, C.~Nobre, O.~Deussen, M.~Meyer, J.~M. Phillips, and A.~Lex.
\newblock Predicting intent behind selections in scatterplot visualizations.
\newblock {\em Information Visualization}, 20(4):207--228, 2021. doi: {{%
10\hspace{.1pt}\discretionary{.}{%
}{.}\hspace{.4pt}1177\discretionary{/}{%
}{/}14738716211038604}}


\bibitem{guo2023datamator}
Y.~Guo, N.~Cao, L.~Cai, Y.~Wu, D.~Weiskopf, D.~Shi, and Q.~Chen.
\newblock Datamator: {A}n authoring tool for creating datamations via data query decomposition.
\newblock {\em Applied Sciences}, 13(17):9709, 2023. doi: {{%
10\hspace{.1pt}\discretionary{.}{%
}{.}\hspace{.4pt}3390\discretionary{/}{%
}{/}app13179709}}


\bibitem{hajiabolhassani2013computational}
A.~Haji-Abolhassani and J.~J. Clark.
\newblock A computational model for task inference in visual search.
\newblock {\em Journal of Visualization}, 13(3):29--29, 2013. doi: {{%
10\hspace{.1pt}\discretionary{.}{%
}{.}\hspace{.4pt}1167\discretionary{/}{%
}{/}13\hspace{.1pt}\discretionary{.}{%
}{.}\hspace{.4pt}3\hspace{.1pt}\discretionary{.}{%
}{.}\hspace{.4pt}29}}


\bibitem{hart1988tlx}
S.~G. Hart and L.~E. Staveland.
\newblock Development of {NASA-TLX} (task load index): {R}esults of empirical and theoretical research.
\newblock In P.~A. Hancock and N.~Meshkati, eds., {\em Human Mental Workload}, vol.~52 of {\em Advances in Psychology}, pp. 139--183. North-Holland, Amsterdam, 1988. doi: {{%
10\hspace{.1pt}\discretionary{.}{%
}{.}\hspace{.4pt}1016\discretionary{/}{%
}{/}S0166\discretionary{%
}{-}{-}4115\discretionary{%
}{(}{(}08\discretionary{)}{%
}{)}62386\discretionary{%
}{-}{-}9}}


\bibitem{Heer2007}
J.~Heer and G.~G. Robertson.
\newblock Animated transitions in statistical data graphics.
\newblock {\em {IEEE} Transactions on Visualization and Computer Graphics}, 13(6):1240--1247, 2007. doi: {{%
10\hspace{.1pt}\discretionary{.}{%
}{.}\hspace{.4pt}1109\discretionary{/}{%
}{/}TVCG\hspace{.1pt}\discretionary{.}{%
}{.}\hspace{.4pt}2007\hspace{.1pt}\discretionary{.}{%
}{.}\hspace{.4pt}70539}}


\bibitem{jolliffe2002principal}
I.~T. Jolliffe.
\newblock {\em {Principal Component Analysis}}.
\newblock Springer-Verlag, New York, 2002. doi: {{%
10\hspace{.1pt}\discretionary{.}{%
}{.}\hspace{.4pt}1007\discretionary{/}{%
}{/}b98835}}


\bibitem{kim2019designing}
Y.~Kim, M.~Correll, and J.~Heer.
\newblock Designing animated transitions to convey aggregate operations.
\newblock {\em Computer Graphics Forum}, 38(3):541--551, 2019. doi: {{%
10\hspace{.1pt}\discretionary{.}{%
}{.}\hspace{.4pt}1111\discretionary{/}{%
}{/}cgf\hspace{.1pt}\discretionary{.}{%
}{.}\hspace{.4pt}13709}}


\bibitem{Kriglstein2012}
S.~Kriglstein, M.~Pohl, and C.~Stachl.
\newblock Animation for time-oriented data: an overview of empirical research.
\newblock In {\em Proceedings of 16th International Conference on Information Visualisation}, pp. 30--35. {IEEE}, 2012. doi: {{%
10\hspace{.1pt}\discretionary{.}{%
}{.}\hspace{.4pt}1109\discretionary{/}{%
}{/}IV\hspace{.1pt}\discretionary{.}{%
}{.}\hspace{.4pt}2012\hspace{.1pt}\discretionary{.}{%
}{.}\hspace{.4pt}16}}


\bibitem{Mcilreavy2019}
L.~Mcilreavy, T.~C.~A. Freeman, and J.~T. Erichsen.
\newblock Two-dimensional analysis of smooth pursuit eye movements reveals quantitative deficits in precision and accuracy.
\newblock {\em Translational Vision Science \& Technology}, 8(5):7--7, 2019. doi: {{%
10\hspace{.1pt}\discretionary{.}{%
}{.}\hspace{.4pt}1167\discretionary{/}{%
}{/}tvst\hspace{.1pt}\discretionary{.}{%
}{.}\hspace{.4pt}8\hspace{.1pt}\discretionary{.}{%
}{.}\hspace{.4pt}5\hspace{.1pt}\discretionary{.}{%
}{.}\hspace{.4pt}7}}


\bibitem{mcinnes2018umap}
L.~McInnes, J.~Healy, and J.~Melville.
\newblock {UMAP}: {U}niform manifold approximation and projection for dimension reduction, 2018. doi: {{%
10\hspace{.1pt}\discretionary{.}{%
}{.}\hspace{.4pt}48550\discretionary{/}{%
}{/}arXiv\hspace{.1pt}\discretionary{.}{%
}{.}\hspace{.4pt}1802\hspace{.1pt}\discretionary{.}{%
}{.}\hspace{.4pt}03426}}


\bibitem{Miller1956}
G.~A. Miller.
\newblock Human memory and the storage of information.
\newblock {\em {IRE} Transactions on Information Theory}, 2(3):129--137, 1956. doi: {{%
10\hspace{.1pt}\discretionary{.}{%
}{.}\hspace{.4pt}1109\discretionary{/}{%
}{/}TIT\hspace{.1pt}\discretionary{.}{%
}{.}\hspace{.4pt}1956\hspace{.1pt}\discretionary{.}{%
}{.}\hspace{.4pt}1056815}}


\bibitem{Nakakoji2001}
K.~Nakakoji, A.~Takashima, and Y.~Yamamoto.
\newblock Cognitive effects of animated visualization in exploratory visual data analysis.
\newblock In {\em Proceedings of International Conference on Information Visualisation}, pp. 77--84. {IEEE} Computer Society, 2001. doi: {{%
10\hspace{.1pt}\discretionary{.}{%
}{.}\hspace{.4pt}1109\discretionary{/}{%
}{/}IV\hspace{.1pt}\discretionary{.}{%
}{.}\hspace{.4pt}2001\hspace{.1pt}\discretionary{.}{%
}{.}\hspace{.4pt}942042}}


\bibitem{nguyen2020evaluation}
Q.~V. Nguyen, N.~Miller, D.~Arness, W.~Huang, M.~L. Huang, and S.~Simoff.
\newblock Evaluation on interactive visualization data with scatterplots.
\newblock {\em Visual Informatics}, 4(4):1--10, 2020. doi: {{%
10\hspace{.1pt}\discretionary{.}{%
}{.}\hspace{.4pt}1016\discretionary{/}{%
}{/}j\hspace{.1pt}\discretionary{.}{%
}{.}\hspace{.4pt}visinf\hspace{.1pt}\discretionary{.}{%
}{.}\hspace{.4pt}2020\hspace{.1pt}\discretionary{.}{%
}{.}\hspace{.4pt}09\hspace{.1pt}\discretionary{.}{%
}{.}\hspace{.4pt}004}}


\bibitem{perin2018assessing}
C.~Perin, T.~Wun, R.~Pusch, and S.~Carpendale.
\newblock Assessing the graphical perception of time and speed on {2D}+time trajectories.
\newblock {\em {IEEE} Transactions on Visualization and Computer Graphics}, 24(1):698--708, 2018. doi: {{%
10\hspace{.1pt}\discretionary{.}{%
}{.}\hspace{.4pt}1109\discretionary{/}{%
}{/}tvcg\hspace{.1pt}\discretionary{.}{%
}{.}\hspace{.4pt}2017\hspace{.1pt}\discretionary{.}{%
}{.}\hspace{.4pt}2743918}}


\bibitem{prolific}
{Prolific Academic Ltd.}
\newblock Prolific -- {Q}uickly find research participants you can trust.
\newblock \url{https://www.prolific.co/}, accessed 2023-09-16.

\bibitem{Purves2001}
D.~Purves, G.~J. Augustine, D.~Fitzpatrick, L.~C. Katz, A.-S. LaMantia, J.~O. McNamara, and S.~M. Williams., eds.
\newblock {\em Neuroscience}.
\newblock Sinauer Associates, Sunderland, 2nd ed., 2001.

\bibitem{quinlan1993combining}
J.~R. Quinlan.
\newblock Combining instance-based and model-based learning.
\newblock In P.~E. Utgoff, ed., {\em Proceedings of 10th International Conference on Machine Learning}, pp. 236--243. Morgan Kaufmann, 1993. doi: {{%
10\hspace{.1pt}\discretionary{.}{%
}{.}\hspace{.4pt}1016\discretionary{/}{%
}{/}B978\discretionary{%
}{-}{-}1\discretionary{%
}{-}{-}55860\discretionary{%
}{-}{-}307\discretionary{%
}{-}{-}3\hspace{.1pt}\discretionary{.}{%
}{.}\hspace{.4pt}50037\discretionary{%
}{-}{-}X}}


\bibitem{Robertson2008}
G.~G. Robertson, R.~Fernandez, D.~Fisher, B.~Lee, and J.~T. Stasko.
\newblock Effectiveness of animation in trend visualization.
\newblock {\em {IEEE} Transactions on Visualization and Computer Graphics}, 14(6):1325--1332, 2008. doi: {{%
10\hspace{.1pt}\discretionary{.}{%
}{.}\hspace{.4pt}1109\discretionary{/}{%
}{/}TVCG\hspace{.1pt}\discretionary{.}{%
}{.}\hspace{.4pt}2008\hspace{.1pt}\discretionary{.}{%
}{.}\hspace{.4pt}125}}


\bibitem{robinson1965mechanics}
D.~A. Robinson.
\newblock The mechanics of human smooth pursuit eye movement.
\newblock {\em The Journal of Physiology}, 180(3):569--591, 1965. doi: {{%
10\hspace{.1pt}\discretionary{.}{%
}{.}\hspace{.4pt}1113\discretionary{/}{%
}{/}jphysiol\hspace{.1pt}\discretionary{.}{%
}{.}\hspace{.4pt}1965\hspace{.1pt}\discretionary{.}{%
}{.}\hspace{.4pt}sp007718}}


\bibitem{rodrigues2023interactive}
N.~Rodrigues.
\newblock Interactive demo of study: Evaluation of animated scatter plot transitions.
\newblock \url{https://github.com/NilsRodrigues/animated-scatterplot-transitions-for-comparative-study}, accessed 2024-01-04, 2024.

\bibitem{rodrigues2022evaluation}
N.~Rodrigues, F.~Dennig, V.~Brandt, D.~Keim, and D.~Weiskopf.
\newblock Evaluation of animated scatter plot transitions, 2022.
\newblock OSF Preregistration. doi: {{%
10\hspace{.1pt}\discretionary{.}{%
}{.}\hspace{.4pt}17605\discretionary{/}{%
}{/}OSF\hspace{.1pt}\discretionary{.}{%
}{.}\hspace{.4pt}IO\discretionary{/}{%
}{/}3E6G4}}


\bibitem{darusAnim}
N.~Rodrigues, F.~L. Dennig, V.~Brandt, D.~A. Keim, and D.~Weiskopf.
\newblock Supplemental material for comparative evaluation of animated scatter plot transitions, 2024.
\newblock DaRUS. doi: {{%
10\hspace{.1pt}\discretionary{.}{%
}{.}\hspace{.4pt}18419\discretionary{/}{%
}{/}darus\discretionary{%
}{-}{-}3451}}


\bibitem{rodrigues2020cluster}
N.~Rodrigues, C.~Schulz, A.~Lhuillier, and D.~Weiskopf.
\newblock Cluster-flow parallel coordinates: tracing clusters across subspaces.
\newblock In {\em Graphics Interface Conference}, pp. 382--392. Canadian Human-Computer Communications Society, 2020. doi: {{%
10\hspace{.1pt}\discretionary{.}{%
}{.}\hspace{.4pt}20380\discretionary{/}{%
}{/}GI2020\hspace{.1pt}\discretionary{.}{%
}{.}\hspace{.4pt}38}}


\bibitem{rottach1996comparison}
K.~G. Rottach, A.~Z. Zivotofsky, V.~E. Das, L.~Averbuch-Heller, A.~O. Discenna, A.~Poonyathalang, and R.~J. Leigh.
\newblock Comparison of horizontal, vertical and diagonal smooth pursuit eye movements in normal human subjects.
\newblock {\em Vision Research}, 36(14):2189--2195, 1996. doi: {{%
10\hspace{.1pt}\discretionary{.}{%
}{.}\hspace{.4pt}1016\discretionary{/}{%
}{/}0042\discretionary{%
}{-}{-}6989\discretionary{%
}{(}{(}95\discretionary{)}{%
}{)}00302\discretionary{%
}{-}{-}9}}


\bibitem{sanftmann20123d}
H.~Sanftmann and D.~Weiskopf.
\newblock {3D} scatterplot navigation.
\newblock {\em {IEEE} Transactions on Visualization and Computer Graphics}, 18(11):1969--1978, 2012. doi: {{%
10\hspace{.1pt}\discretionary{.}{%
}{.}\hspace{.4pt}1109\discretionary{/}{%
}{/}TVCG\hspace{.1pt}\discretionary{.}{%
}{.}\hspace{.4pt}2012\hspace{.1pt}\discretionary{.}{%
}{.}\hspace{.4pt}35}}


\bibitem{schulz2016generative}
C.~Schulz, A.~Nocaj, M.~El{-}Assady, S.~Frey, M.~Hlawatsch, M.~Hund, G.~K. Karch, R.~Netzel, C.~Sch{\"{a}}tzle, M.~Butt, D.~A. Keim, T.~Ertl, U.~Brandes, and D.~Weiskopf.
\newblock Generative data models for validation and evaluation of visualization techniques.
\newblock In {\em Proceedings of the Sixth Workshop on Beyond Time and Errors on Novel Evaluation Methods for Visualization}, pp. 112--124. {ACM}, 2016. doi: {{%
10\hspace{.1pt}\discretionary{.}{%
}{.}\hspace{.4pt}1145\discretionary{/}{%
}{/}2993901\hspace{.1pt}\discretionary{.}{%
}{.}\hspace{.4pt}2993907}}


\bibitem{gapminder}
{Stiftelsen Gapminder}.
\newblock {Gapminder Tools}.
\newblock \url{https://www.gapminder.org/tools/\#\$chart-type=bubbles\&url=v1}, accessed 2023-09-16.

\bibitem{taylor2010ceiling}
T.~H. Taylor.
\newblock Ceiling effect.
\newblock In N.~J. Salkind, ed., {\em Encyclopedia of Research Design}, pp. 133--134. SAGE Publications, Inc., Thousand Oaks, 2010. doi: {{%
10\hspace{.1pt}\discretionary{.}{%
}{.}\hspace{.4pt}4135\discretionary{/}{%
}{/}9781412961288}}


\bibitem{Tekusova2007}
T.~Tekušová and J.~Kohlhammer.
\newblock Applying animation to the visual analysis of financial time-dependent data.
\newblock In {\em Proceedings of 11th International Conference Information Visualization}, pp. 101--108. {IEEE}, 2007. doi: {{%
10\hspace{.1pt}\discretionary{.}{%
}{.}\hspace{.4pt}1109\discretionary{/}{%
}{/}iv\hspace{.1pt}\discretionary{.}{%
}{.}\hspace{.4pt}2007\hspace{.1pt}\discretionary{.}{%
}{.}\hspace{.4pt}28}}


\bibitem{thompson2021dataanimator}
J.~R. Thompson, Z.~Liu, and J.~Stasko.
\newblock Data animator: {A}uthoring expressive animated data graphics.
\newblock In {\em Proceedings of the {CHI} Conference on Human Factors in Computing Systems}. {ACM}, 2021. doi: {{%
10\hspace{.1pt}\discretionary{.}{%
}{.}\hspace{.4pt}1145\discretionary{/}{%
}{/}3411764\hspace{.1pt}\discretionary{.}{%
}{.}\hspace{.4pt}3445747}}


\bibitem{tominski2021flexible}
C.~Tominski, G.~Andrienko, N.~Andrienko, S.~Bleisch, S.~I. Fabrikant, E.~Mayr, S.~Miksch, M.~Pohl, and A.~Skupin.
\newblock Toward flexible visual analytics augmented through smooth display transitions.
\newblock {\em Visual Informatics}, 5(3):28--38, 2021. doi: {{%
10\hspace{.1pt}\discretionary{.}{%
}{.}\hspace{.4pt}1016\discretionary{/}{%
}{/}j\hspace{.1pt}\discretionary{.}{%
}{.}\hspace{.4pt}visinf\hspace{.1pt}\discretionary{.}{%
}{.}\hspace{.4pt}2021\hspace{.1pt}\discretionary{.}{%
}{.}\hspace{.4pt}06\hspace{.1pt}\discretionary{.}{%
}{.}\hspace{.4pt}004}}


\bibitem{Utts2005}
J.~M. Utts.
\newblock {\em {S}eeing {T}hrough {S}tatistics}.
\newblock Thomson Brooks/Cole, Belmont, 3rd ed., 2005.

\bibitem{wang2018vector}
Y.~Wang, D.~Archambault, C.~E. Scheidegger, and H.~Qu.
\newblock A vector field design approach to animated transitions.
\newblock {\em {IEEE} Transactions on Visualization and Computer Graphics}, 24(9):2487--2500, 2018. doi: {{%
10\hspace{.1pt}\discretionary{.}{%
}{.}\hspace{.4pt}1109\discretionary{/}{%
}{/}TVCG\hspace{.1pt}\discretionary{.}{%
}{.}\hspace{.4pt}2017\hspace{.1pt}\discretionary{.}{%
}{.}\hspace{.4pt}2750689}}


\bibitem{yao2022visualization}
L.~Yao, A.~Bezerianos, R.~Vuillemot, and P.~Isenberg.
\newblock Visualization in motion: A research agenda and two evaluations.
\newblock {\em {IEEE} Transactions on Visualization and Computer Graphics}, 28(10):3546--3562, 2022. doi: {{%
10\hspace{.1pt}\discretionary{.}{%
}{.}\hspace{.4pt}1109\discretionary{/}{%
}{/}tvcg\hspace{.1pt}\discretionary{.}{%
}{.}\hspace{.4pt}2022\hspace{.1pt}\discretionary{.}{%
}{.}\hspace{.4pt}3184993}}


\bibitem{yee2001animated}
K.~Yee, D.~Fisher, R.~Dhamija, and M.~A. Hearst.
\newblock Animated exploration of dynamic graphs with radial layout.
\newblock In {\em {IEEE} Symposium on Information Visualization}, pp. 43--50. {IEEE} Computer Society, 2001. doi: {{%
10\hspace{.1pt}\discretionary{.}{%
}{.}\hspace{.4pt}1109\discretionary{/}{%
}{/}INFVIS\hspace{.1pt}\discretionary{.}{%
}{.}\hspace{.4pt}2001\hspace{.1pt}\discretionary{.}{%
}{.}\hspace{.4pt}963279}}


\end{thebibliography}

\end{document}